\documentclass[usenatbib]{mn2e}
\usepackage{graphicx}
\usepackage{natbib}
\usepackage{amsmath}
\usepackage{amsfonts}
\usepackage{dsfont}
\usepackage{multicol}
\usepackage{widetext}
\usepackage{color}

\newcommand\noise{{\cal N}}

\def\apjl{{ApJ}}                % Astrophysical Journal, Letters
\def\apjs{{ApJS}}               % Astrophysical Journal, Supplement
\def\ao{{Appl.~Opt.}}           % Applied Optics
\def\aap{{A\&A}}                % Astronomy and Astrophysics
\begin{document}
 
\title[Lensing $+$ redshift surveys]{Combining weak lensing tomography and spectroscopic redshift surveys} 
\author[Cai \& Bernstein]{Yan-Chuan Cai\thanks{E-mail: yancai@sas.upenn.edu} and Gary Bernstein\\
 Department of Physics and Astronomy, University of Pennsylvania, 
 Philadelphia, PA 19104\\
 Center for Particle Cosmology, University of Pennsylvania, 
 Philadelphia, PA 19104}
\maketitle
\begin{abstract}
Redshift space distortion (RSD) is a powerful way of measuring the
growth of structure and testing General Relativity, but it is limited by
cosmic variance and the degeneracy between galaxy bias $b$ and the growth rate factor $f$. 
The cross-correlation of lensing shear with the galaxy density field can
in principle measure $b$ in a manner free from cosmic variance limits, breaking the $f-b$ degeneracy
and allowing inference of
the matter power spectrum from the galaxy survey. We analyze the
growth constraints from a realistic tomographic weak lensing photo-$z$ survey
combined with a spectroscopic galaxy redshift
survey {\em over the same sky area}. 
For sky coverage $f_{\rm sky}=0.5$, analysis of the transverse modes
measures $b$ to 2--3\% accuracy per 
$\Delta z=0.1$ bin at
$z<1$ when $\sim 10$ galaxies arcmin$^{-2}$ are measured in the
lensing survey and all halos with $M>M_{\rm min}=10^{13}h^{-1}M_\odot$ have spectra.
For the gravitational growth parameter
parameter $\gamma$ ($f=\Omega_m^{\gamma})$, combining the lensing
information with RSD analysis of non-transverse modes yields accuracy
$\sigma(\gamma)\approx0.01$. Adding lensing information to the RSD
survey improves $\sigma(\gamma)$ by an amount equivalent to a
$3\times$ ($10\times$) increase in RSD survey area when the
spectroscopic survey extends down to halo mass $10^{13.5}$ ($10^{14}$)
$h^{-1} M_\odot$.  
We also find that the $\sigma(\gamma)$ of overlapping surveys is
equivalent to that of surveys 1.5--2$\times$ larger if they are
separated on the sky.  This gain is greatest when the spectroscopic
mass threshold 
is $10^{13}$--$10^{14}h^{-1} M_\odot$, similar to LRG surveys.  The
gain of overlapping surveys is reduced for very deep or very shallow
spectroscopic surveys, but any practical surveys are more powerful
when overlapped than when separated.  The gain of overlapped surveys
is larger in the case when the primordial power spectrum normalization
is uncertain by $>0.5\%$.
\end{abstract}

\section{Introduction}
Measurement of the linear growth of structure of the Universe is essential in that the growth history reflects 
the nature of dark energy and the underlying gravity model \citep[e.g.][]{Yamamoto2010}, i.e. whether dark energy 
is a cosmological constant, or is evolving with time, or if General
Relativity (GR) is the correct gravity model 
that governs the evolution of the Universe. In the linear regime of GR, the growth of perturbations is scale independent. It 
can be parameterized as the linear growth function $G$, with
$P(z)=G^2(z)P_{\rm CMB}$, where $P(z)$ and $P_{\rm CMB}$ are the matter density power spectra at redshift $z$ and at the epoch of recombination, respectively. $G(z)$ carries information about 
the amount of dark energy and dark matter. 
The growth rate factor $f \equiv \frac{\partial \ln G}{\partial \ln a}$, with $a$ being the scale factor, 
is another quantify of interest: $f$ can be well approximated as $f=\Omega_m^{\gamma}$, with $\gamma$ in 
a narrow range near 0.55,  for a wide variety of dark-energy models in General Relativity 
\citep{Peebles80,Lahav91,LinderCahn}.  A precise measure of $\gamma$ therefore 
enables one to distinguish GR from alternative gravity models. In a
braneworld type of modified gravity, for example, $\gamma$ is 
different from $\gamma_{\rm GR}$ by more than $20\%$ \citep{LinderCahn}. 

Precise measurements of $G$ and $f$ constrain dark energy and gravity, and Redshift Space Distortion (RSD) has been 
shown to be a powerful approach to perform this measurement \citep[e.g.][]{Kaiser87, Cole94, 
Hamilton00, Peacock01,  Scoccimarro04, Guzzo08, Cabre09,
Blake11}. This RSD measurement is, however, only 
precise in the linear regime.  At late epochs of the 
Universe, the linear regime (of the velocity field in particular,) 
is confined to very large scales, $k\le 0.1h^{-1}{\rm Mpc}^{-1}$. On these scales, the measurement is usually 
limited by sample variance, or cosmic variance---we do not have many
independent perturbation modes for the measurement because of the
finite survey volume observable in a given epoch.

Using multiple tracers of the density field, one can in principle evade sampling variance, and measure the 
linear growth of structure with unbounded accuracy \citep{MS09,WSP,Gil-Marin10,BC2011}. 
The great benefit of multiple tracers is not realized, however, if
only the clustering of galaxies is measured because of 
the following: using RSD, one can only measure 
$\beta=f/b$ and the product $fG$. Without any prior knowledge of
the galaxy bias, one can not constrain $f$ or $G$ independently.
It has been shown by \citet{BC2011}[BC11] that prior knowledge on galaxy bias significantly
improves the constraint on the growth of structure in the case of
single survey redshift bin.

In principle, galaxy bias can be measured by cross-correlating weak gravitational
lensing convergence with galaxy clustering 
\citep{Pen04}. This bias
measurement is free of sample variance in the sense that the bias
errors from a survey of a fixed number of modes can be reduced without limit as the
lensing measurement noise and galaxy shot noise are decreased.
This enable us to use large-scale 
modes that are well in the linear regime for high-accuracy measurement (see BC11).
Combining a weak lensing survey (using photometric redshifts for
source galaxies) with a spectroscopic redshift survey of lens galaxies
can serve this purpose perfectly: the bias of the spectroscopic
galaxies is measured by cross-correlation with the lensing signal in
transverse modes, while the RSD analysis of the
spectroscopic sample is conducted using non-transverse modes over the same volume.

Current and future large surveys are making possible the combination
of spectroscopic redshift-space and lensing maps over 
common volumes. For example, the footprints of the upcoming {\em Dark
  Energy Survey (DES)}\footnote{http://www.darkenergysurvey.org} may   
overlap with that of an extended {\em Baryon Oscillation Spectroscopic
  Survey (BOSS)}\footnote{http://cosmology.lbl.gov/BOSS} survey near the  
equator. The future {\em Euclid} space telescope \citep{EUCLID} 
is designed to take spectra and images of galaxies at the same time. 

In this work, we will explore 
the potential improvements in constraint of growth of structure and
gravity from overlapping RSD and lensing surveys.
This is an extension of BC11, where we consider 
the case of one single RSD redshift bin with an arbitrarily assigned prior on galaxy bias. In this work, we will consider 
the more realistic case of tomography using spectroscopic (RSD) and
photometric (lensing) surveys covering common sky area, with
both types of tracers divided into as many as 20 redshift bins. We will 
explore how well galaxy bias can be measured using the cross-correlation of galaxy shear 
and galaxy clustering in this realistic joint tomographic survey. The basic scheme is:
\begin{enumerate}
\item Conduct a galaxy redshift survey and a weak lensing (photo-$z$) survey over the same volume of the Universe. 
Split both galaxy samples into redshift bins. 

\item Optimally weight galaxies in each bin of the redshift survey to produce
  a mass density estimator with minimal stochasticity
  \citep{Hamaus10,CBS11}.  Measure the 2-point shear-shear (from the
  lensing survey), and density-density (from the spectroscopic survey)
  correlations and the shear-density cross correlations between all
  $z$-bins.  Using these measurements of the covariance in transverse
  modes, constrain the bias $b$ of the spectroscopic galaxy density
  estimator and the mass power spectrum $P_m$ in each redshift bin.
\label{step2}
\item In each $z$-bin, split galaxies from the redshift survey into different bias bins.
Perform multiple-tracer RSD measurement \citep{MS09, BC2011} using the
redshift-space density field of the binned galaxies. The $b$ and 
$P_m$ constraints derived from transverse modes in step \ref{step2} are
incorporated to break the $f-b$ degeneracy inherent to RSD, so 
that separate constraints on $G$ and $f$ can be achieved. Throughout the paper, 
we will suppress the latin index denoting redshift in equations that involve only a single redshift bin, 
such as the RSD Fisher matrix. We use Greek indices for galaxy bias bins.
\end{enumerate}

We use the Fisher matrix method to forecast growth constraints
resulting from a model survey consisting of spectroscopic and
lensing surveys covering a common $f_{\rm sky}=0.5$ of the sky, reaching the 
depth of $z=2$.  We split both samples into 20 $z$-bins of width $\Delta z=0.1$. 
We employ the halo model for our survey model, assuming that each halo
above mass $M_{\rm min}$ hosts one spectroscopic target galaxy.
We set up our forecast methodology for lensing tomography in section \ref{lensing}
and for multi-tracer RSD in section \ref{RSD}. We 
summarize our numerical results in section \ref{result}, and conclude in section \ref{conclusion}.

Unless noted otherwise, we assume a fiducial flat $\Lambda$CDM cosmology with the following parameters: 
$\Omega_m$=0.272, $\Omega_{\Lambda}=0.728$, $\Omega_b=0.0455$,
$\sigma_8=0.807$, $n_s=0.961$, $H_0=70.2$. \citep{WMAP7}

\section{galaxy-shear cross-correlation}
\label{lensing}
\subsection{Weak lensing tomography}
Weak gravitational lensing of background galaxies is a powerful way to measure the projected mass 
density of the foreground. It is free from galaxy bias and can be used to measure galaxy bias when 
cross-correlating with the galaxy density field. Source galaxies split into different tomographic bins enable 
us to probe the mass density at different epochs of the Universe. 

For the $i$th $z$-bin of source galaxies, the observable of weak lensing is the distortion of those 
galaxy images, or cosmic shear, which is induced by foreground large-scale gravitational potential. 
From the cosmic shear one can infer the convergence $\kappa$, which is
a weighted projection of the 3-D mass density of the foreground: 
\begin{equation}
\kappa_i(\theta)=\frac{3H_0^2\Omega_m}{2c^2}\int^{\chi_i}_{0}\chi W_i(\chi)\frac{\delta(\chi\theta, \chi)}{a(\chi)}d\chi
\end{equation}
where $\chi$ is the comoving radial distance, $a$ is the scale factor of the Universe, $\delta$ is the 3-D matter 
density contrast.  The lensing weight function is
\begin{equation}
W_i(\chi)=\frac{1}{\bar n_i}\int^{\chi_{\rm H}}_{0} n_i(z)\frac{dz}{d\chi_s}\frac{\chi_s-\chi}{\chi_s}d\chi_s,
\end{equation}
where $\bar n_i$ is the number of galaxies in the $i$th redshift bin,
distributed as $n_i(z)$. $\chi_{\rm H}$ is the horizon distance.
We assume a total source redshift distribution of the form
\begin{equation}
n(z)=n_0z^2\exp(-z/z_0),  %*(180.0*60.0/!PI)^2*(2.0*!PI),
\end{equation}
where $z_0=0.45$ is chosen to fit with the predicted Euclid survey's median redshift, 
and $n_0$ is chosen such that $\int n(z) dz=N_{\rm lens}$, the total density of
lens source galaxies per steradian. 
In section \ref{result}, we will examine results for a wide range of $N_{\rm lens}$.

The weak lensing signal is detectable only in statistics of large
source-galaxy ensembles, e.g. via the
two-point correlation function or its Fourier space counterpart, the shear 
power spectrum, or higher order correlation functions. \citet{Bernstein09} 
gives a framework for two-point analysis of weak lensing survey data. 
We follow the notation of \citet{Bernstein09} for the Fisher matrix
from lensing tomography. 
We will work in the Fourier domain, and employ the Limber
approximation \citep[e.g.][]{Limber54, Kaiser92, Hu00,Verde00, Cai09},  
assuming that there is no correlation between $\delta$ in different redshift bins nor between 
different spherical harmonics. We also assume that within each redshift bin, $n_i(z)$ is 
a Dirac delta function at $z_i$, and that the projected mass
fluctuations $\delta_i$ within bin $i$ can be treated as a single lens
deflection screen at $z_i$. Under these assumptions, for a given spherical harmonic, 
the convergence of the $i$th source galaxy bin is just the weighted sum of the mass 
density of all the redshift bins in front of the $i$th bin ($i$ increases with redshift):
\begin{equation}
\label{kappa}
\kappa_i(l)=\sum_{k=1}^{i-1}A_{ik}\Delta\chi_k\delta_k(l)F_k+ \epsilon_i,
\end{equation}
with $\langle \epsilon_i \epsilon_j \rangle = \delta^K_{ij} \sigma_{\epsilon}^2/ n_i$ the 
variance of lensing shear noise.  We take $ \sigma_{\epsilon}=0.22$ throughout our calculation.
$A_{ik}=\frac{D_i-D_k}{D_i}$, $D_i$ is the comoving angular diameter
distance to redshift $z_i$, and $F_k=\frac{3}{2}H_0^2\Omega_m(1+z_k)$. 

\subsection{Covariance matrix for lensing and galaxy density}
When combining lensing tomography with a galaxy redshift survey over
the same volume, we assume that the spectroscopic galaxies will be
split into $z$ bins matching the source bins.  A projected
density estimator $\delta_g$ will be produced in each bin using some
weighted combination of the spectroscopic galaxies.
These projected density estimates are essentially the 
transverse modes of the RSD measurement in section \ref{RSD}. Each
galaxy is given an optimal weight as described
in section \ref{Stochasticity}.  
The spectroscopic galaxies can have a different selection function from
the lensing source galaxies.  In cases where spectroscopic galaxies
are not available (such as when we consider non-overlapping RSD and
lensing surveys), we will assume that the $\delta_g$ measurement is
made using galaxies with photometric redshift assignments from the
lensing survey's imaging data.

These measurements will be made for each mode transverse to the
line of sight, indexed by spherical harmonic $l$:
\begin{enumerate}
\item $C^{\kappa \kappa}_{ij}(l)$---the (cross-) power spectrum of the lensing convergence at (and between) different 
redshift slices from the lensing survey. 

\item $C^{g g}_{ij}(l)$--- the power spectra at each
redshift slice of the projected galaxy density estimator formed from the
weighted spectroscopic galaxy survey (or photo-$z$ sample).
We assume no correlation between densities of distinct redshift
slices, following the Limber approximation and ignoring magnification
biases and redshift mis-assignments, so ${\bf C}^{gg}$ is diagonal.

\item $C^{g \kappa}_{ij}(l)$---the shear-galaxy cross-spectra between different redshift slices. Galaxy density will only
correlate with shear in the background, so ${\bf C}^{g\kappa}$ is a
triangular matrix.
\end{enumerate}

More specifically, the measurements can be expressed as:
\begin{eqnarray}
C^{\kappa\kappa}_{ij}(l) &=&\sum_{k=1}^{min\{i,j\}-1}A_{ik}A_{jk}\Delta\chi_kP_k(l)F_k^2+\sigma_{\epsilon}^2\delta_{ij}/n_i \nonumber \\
C^{gg}_{ij}(l) &=&D_i^{-2}\Delta\chi_i^{-1}P_i(l/D_i)\bar b_i^2\delta_{ij}+\noise_i(l)  \nonumber \\
C^{g\kappa}_{ij}(l) &=&A_{ji}D_i^{-1}P_i(l/D_i)\bar b_iF_i \ |_{ i<j} 
\end{eqnarray}
where $n_i$ is the number density of lens source galaxies per steradian at the $i$th redshift bin; 
$P_i(k)=G_i^2P_{\rm CMB}(k)$ is the 3-D mass power spectrum at the $i$th redshift slice, with 
$G_i$ being the linear growth function at $z_i$; $l=k/D_i$ is the
angular wavenumber; and $\bar b_i$ and $\noise_i(l)$ are the
scale-independent bias and stochastic noise power, respectively, of the weighted
spectroscopic galaxy density estimator at $z_i$.  Note that we ignore
complications from intrinsic alignments of galaxies, photometric
redshift errors, and other lensing measurement systematic errors.

The fiducial value of the noise term $\noise_i(l)$ is taken from a
fiducial stochasticity $E_i(l)$ of the galaxy density estimator:
$\noise_i^{\rm fid}(l)=E_i(l)^2D_i^{-2}\Delta\chi_i^{-1}P_i(l/D_i)\bar
b_i^2\delta_{ij}$.  Our model for the fiducial
$E_i(l)$ is taken from \citet{CBS11} and described in section \ref{Stochasticity}. 
The fiducial value of $\noise_i(l)$ 
is taken from this model, but $\noise_i(l)$ is still treated as an independent free 
parameter of $C^{gg}_{ii}(l)$, and will be marginalized over.  Note
that most analyses (including our own RSD Fisher matrix) assume that
the stochastic power is known {\it a priori} to be given by the
Poisson formula.  We find in this lensing analysis that a strong prior
knowledge of $\noise$ can substantially influence the final growth
constraints, so we adopt a weak conservative prior quantified below.

For each spherical harmonic $l$, the full covariance matrix for the
lensing and density measurements is
\[ {\bf C}=\left[ \begin{array}{cc}
{\bf C^{\kappa\kappa}} &  {\bf C^{g \kappa}} \\
{\bf (C^{g \kappa})^T} &  {\bf C^{gg}}  \end{array} \right],\]
which is a $40\times40$ matrix for our
20 redshift slices $z_i={0.1, 0.2...2.0}$.

\subsection{Fisher matrix of the cross-correlation}
\label{cross-correlation}
The free parameters of the model for the lensing Fisher matrix are:
\begin{enumerate}
\item The amplitude of the mass power spectrum at different redshift
  $P_i(l/D_i)$, which is in turn a function of only the linear growth
  function $G_i$ at redshift $z_i$, $P_i(l)=G_i^2P_{CMB}(l/D_i)$.  In
  practice we use the parameter $p_i = \ln P_i(l) = 2 \ln G_i + {\rm const}$.

\item The bias $\bar b_i$ of the weighted spectroscopic galaxy
  density. It is related to the biases $b_{i\alpha}$ of individual galaxy bins used
  in the RSD analysis by $\bar b_i = \sum_\alpha w_{i\alpha}
  b_{i\alpha},$ with the weights $w_{i\alpha}$ as assigned in Section \ref{Stochasticity}.
  How galaxies are made into different bias bins is detailed in Section \ref{RSD}

\item The noise $\noise_i(l)$ in the galaxy-galaxy clustering measurement in transverse modes. 
\end{enumerate}
We fix all parameters except for the 20 $P_i$, 20 $\bar b_i$ and 20
$\noise_i$---note that this includes taking the cosmological distances
$D_i$ as known. 
There are no lensing sources behind the highest redshift bin, so the last
bias parameter is unconstrained. 
We therefore drop the rows and columns 
of the Fisher matrix for the $\bar b, P,$ and $\noise$ parameters of
the highest redshift bin, leaving $57\times57$ elements
at each multipole $l$:
\begin{equation}
{\bf F^{p q}_{Lens}}(l)=\frac{1}{2}{\rm Tr}\left[{\bf C}(l)^{-1}{\bf C}(l)_{,p} {\bf C}(l)^{-1} {\bf C}(l)_{,q}\right]
\end{equation}
where $p, q \in \{P_i, \bar b_i, \noise_i\}$. The above equation holds true because $<\kappa>=0$ for 
the whole sky. For the first bin, there is no constraint from lensing, so the first 
row and column of $C_l^{\kappa\kappa}$ are zero. We add priors on the $\noise_i$ parameters to indicate 
uncertainties propotional to the fiducial values:
\begin{equation}
F^{\noise\noise, prior}_{ij}(l)=\delta_{ij} \left(2 \noise_i^{\rm fid} \sqrt{\alpha  N_l}\right)^{-2}. 
\end{equation}
$N_l$ is the 
number of $l$ bins that we use. This scaling produces a prior such
that the mean $\noise_i$ over all $l$ bins in known to accuracy
$2\alpha \noise_i^{\rm fid}$.  We choose the very weak prior
$\alpha=50$ for our calculation.

\begin{figure*}
\begin{center}
\resizebox{\hsize}{!}{
\includegraphics[angle=0]{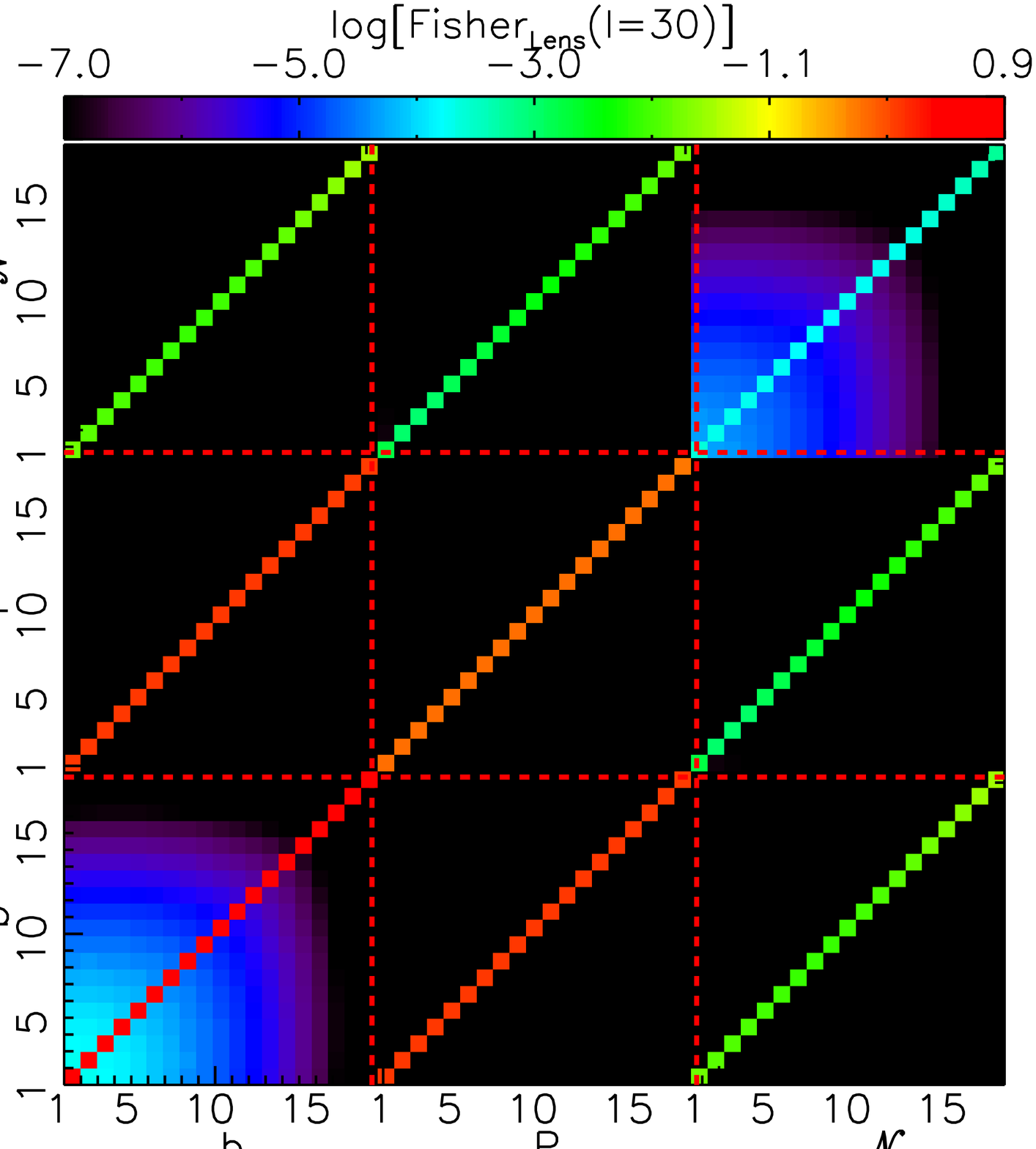}
\includegraphics[angle=0]{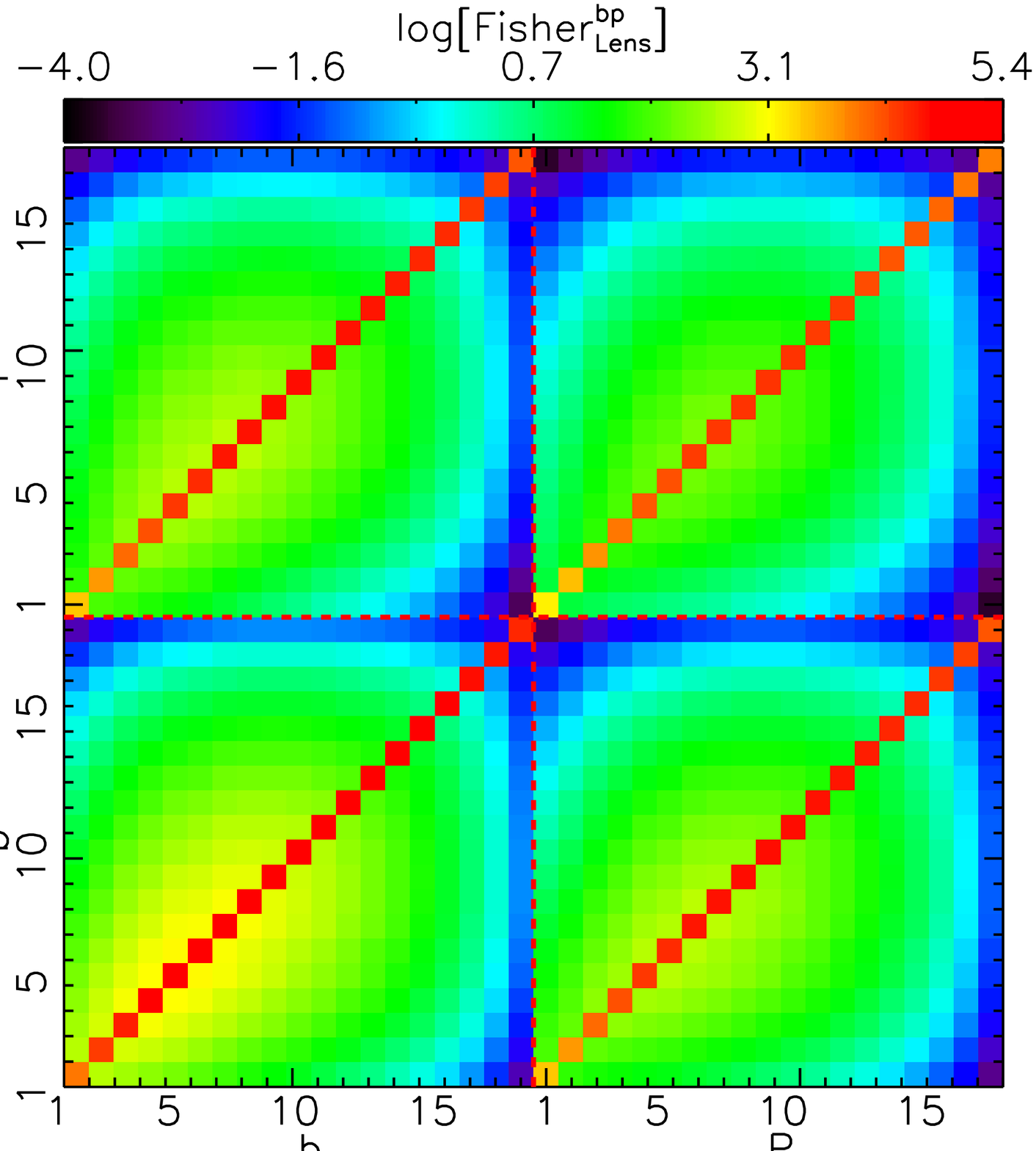}
\includegraphics[angle=0]{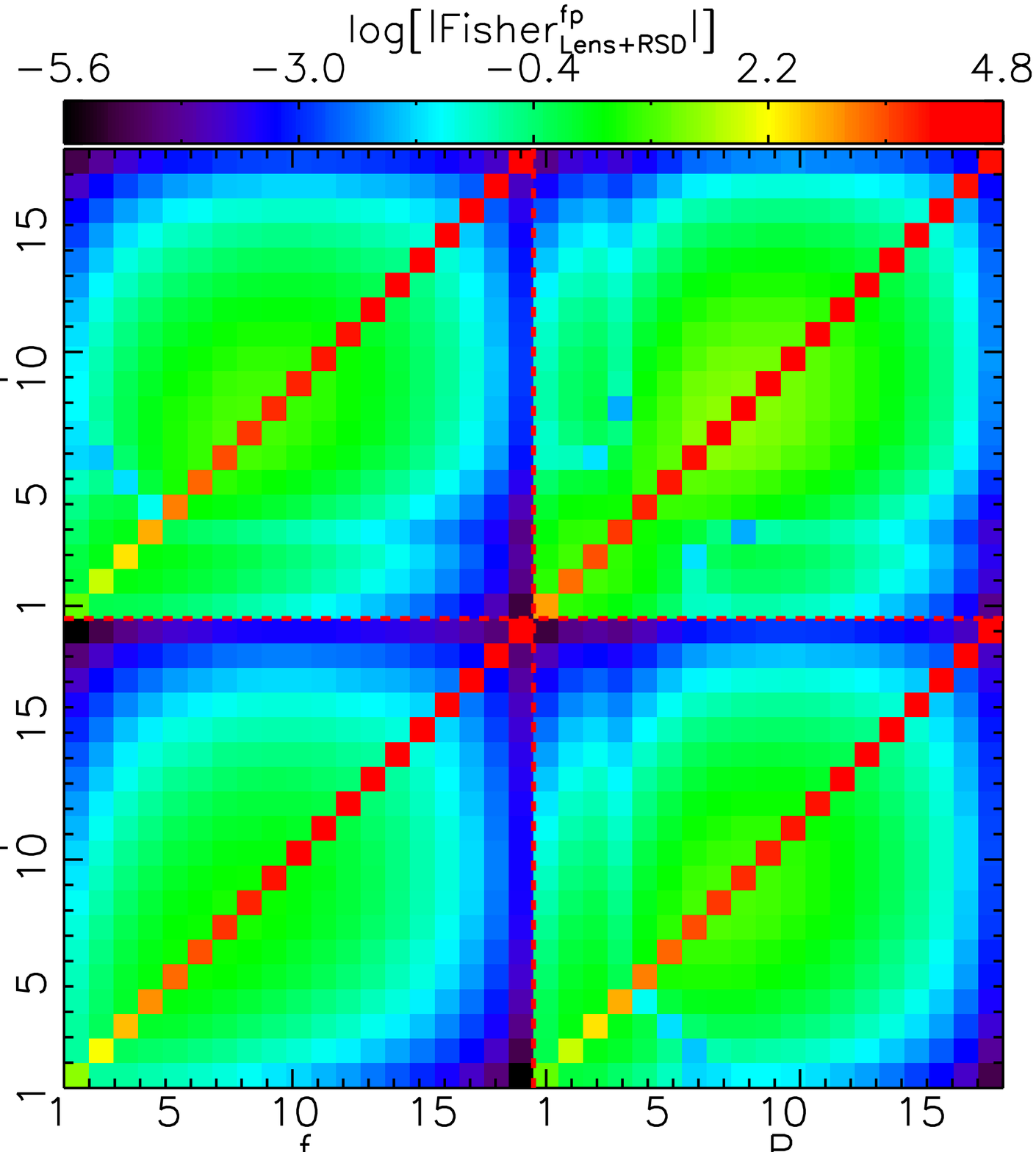}}
\caption{
Examples of Fisher matrices for 20 redshift bins for weak lensing source density $N_{\rm lens}=10{\rm
  arcmin}^{-2}$ and spectrosopic survey depth $M_{\rm
  min}=10^{12}M_{\odot}$; each survey covers $f_{\rm
  sky}=0.5$.  {\it Left:} Fisher matrix ${\bf F_{\rm Lens}}$ from joint lensing/galaxy
measurements on transverse modes at $l=30$, with parameters for bias
$\bar b$, mass power $P$,  and galaxy stochastic power $\noise$ at
each $z$;
{\it Middle:} ${\bf F_{\rm Lens}}$ for linear $\bar b$ and $P$ after marginalizing over all $\bar
b$'s and $P$'s of non-linear modes, summing over all $l$, and 
marginalizing all $\noise$'s;  {\it Right:} Fisher matrix about $f$
and $P$ after summing ${\bf F_{\rm Lens}}+{\bf F_{\rm RSD}}$,
after marginalizing over all biases. 
Each matrix
block contains parameters from low $z$ in lower-left to high $z$ in
upper right, as labeled, omitting the highest $z$ bin which is unconstrained.  
Fisher matrices use logarithms of each parameter so that fractional
errors are represented.
The fiducial value of stochastic powers $\noise$ are estimated from halo mode described in section 
\ref{Stochasticity} and a weak prior is applied.  Note that
correlations between distinct redshifts are always quite weak.
}
\label{FisherbPE}
\end{center}
\end{figure*}

An example of the lensing Fisher matrix is shown in the left panel 
of Figure~\ref{FisherbPE} for a single mode at $l=30$. We find the
Fisher matrix is close to 
block-diagonal, i.e. $\bar b's$, $P's$ and $\noise's$ at distinct $z$ are only weakly correlated. Lowering 
the fiducial stochasticity $\noise_i$ makes the Fisher matrix more diagonal.

 \subsection{Sources of noise in the lensing measurement}
\label{Stochasticity}
In the determination of the $\bar b_i$ using the cross-correlation of
lensing and galaxy surveys, 
both the 
shear measurement noise 
and stochasticity between the tracer and the mass field serve as
sources of error. We can rewrite Equation~(\ref{kappa}) as  
\begin{eqnarray}
\label{kappa2}
\kappa_i(l)&=& \sum_{k=1}^{i-1}A_{ik}\Delta\chi_k F_k\left(\frac{\delta^g_k(l)}{\bar b_k}-e_k(l)\right)+ \epsilon_i, \\
&=& \sum_{k=1}^{i-1}A_{ik}\Delta\chi_kF_k\frac{\delta^g_k(l)}{\bar b_k} - s_i(l) + \epsilon_i,
\end{eqnarray}
with the galaxy overdensity $\delta^g_k(l)$=$\bar b_k\delta_k(l)+e_k(l)$, where
$e_k$ is the stochastic component. 
The lensing observable hence has two
indistinguishable stochastic components that are
{\em not} properly traced by the galaxies---its shear measurement noise
$\epsilon_i$ plus the total convergence from the mass
fluctuations $s_i(l)=\sum_{k=1}^{i-1}A_{ik}\Delta\chi_kF_ke_k(l)$.
Together these degrade the constraint on the mean bias, and also affect the constraint on the mass power since 
$\bar b$ and $P$ are strongly correlated (see the Fisher matrix of $\bar b$ and $P$ at the middle panel of figure \ref{FisherbPE}). 
We will investigate in Section \ref{bpConstraint} how the choices of
lensing source density $N_{\rm lens}$ and spectroscopic depth $M_{\rm min}$, which
set these two noise levels, affect the constraint on $\bar b$ and $P$, and 
further affect constraints on the growth.

Since the stochasticity of the galaxy density has been shown to be a limit for the precision of weak lensing  
constraints on the bias \citep[e.g.][]{Pen04}, we have incentive to
reduce the stochasticity below the commonly assumed Poisson level.  
Sub-Poisson stochasticity has been demonstrated in $N$-body simulations,
\citep[e.g.][]{Bonoli09, Hamaus10, CBS11}. Here we follow the method of \citet{CBS11}, hereafter CBS, for minimizing 
the stochasticity of a mass estimator from a weighted combination of halos.

The optimal weight $w_{\rm opt}$ of each halo is a function of its mass
and of the minimum mass $M_{\rm min}$ of halo included in the survey. The resulting 
stochasticity between the weighted halo field and the mass field $E_{\rm opt}$. 
Explicit expressions from CBS for $w_{\rm opt}$ and $E_{\rm opt}$ are given in the Appendix. 
With this definition of $E_{\rm opt}$, the stochastic components of the galaxy 
density clustering can be written as $<e(l) e'(l)>=E_{\rm opt}^2(l)/[1-E_{\rm opt}^2(l)]$. 
This expression is a function of redshift, but for simplicity, we drop the latin index 
denoting redshift here.

In this work, we use the CBS halo model description of $E_{\rm opt}$
to produce the
fiducial value of stochastic power $\noise_i(l)$. 
In principle, both $w_{\rm opt}$ and $E_{\rm opt}$ 
are functions of the Fourier wave number $k$, the minimal halo mass of the catalogue $M_{\rm min}$ 
and redshift $z$. However, since we find that 
$w_{\rm opt}$ depends very weakly on $k$ in the linear regime, 
we will just adopt the $w_{\rm opt}$ for $k=0.01 h\,{\rm Mpc}^{-1}$ at each $M_{\rm min}$ and $z$.
CBS shows how $E_{\rm opt}$ drops with $M_{\rm min}$; therefore, a deeper spectroscopic 
redshift survey targeting galaxies hosted by lower-mass halos leads to
a higher-precision measure of the galaxy bias when 
cross-correlated with lensing.

In using the CBS model for our fiducial value of $E$, we are assuming
that the host halo mass of each spectroscopic target 
is known by some means, and that the spectroscopic targets are
complete to the limiting halo mass.
This assumption might be somewhat strong but it is not impossible for
a real survey. 
One can imagine using the relatively deep lensing image survey to resolve satellite 
galaxies that are hosted by each spectroscopic galaxy's halo. The
number of satellite galaxies could then be used to 
estimate the halo mass. 
We need a spectroscopic redshift of only the central galaxy of each
halo.

When we are forecasting scenarios in which there is no overlapping
spectroscopic survey for lensing data at a given $z$, we assume that
photometric redshift maps can produce a $\delta_g$ density estimator
with fiducial stochasticity $E_i=0.5$.  Note that the bias $\bar b$
for the photo-$z$ population can {\em not} be used in this case to constrain the
biases of the spectroscopic population because of different selection
functions. 

\subsection{Summation over modes}
\label{nonlin}
In the lensing Fisher matrix, we marginalize over all noise parameters $\noise_i$ to 
have $(19 b's+19 P's)^2$ left. We also marginalized over all those 
b's and P's of non-linear modes. In marginalizing over non-linear modes, we will retain more modes at
high $z$, since for fixed $l$, the physical scale is larger at high
$z$. Furthermore, non-linearity develops on smaller scales at higher redshift.
Since accurate predictions of redshift distortions will likely be
available only in the linear regime, we do not use non-linear modes in our measurements.
%We therefore marginalize 
%over $P$'s and $\bar b$'s at redshifts for which $k=l D_i > k^z_{\rm
%  max}$, with $k^z_{\rm max}$ denoting a maximum linear wavenumber at
%redshift $z$. 

To separate linear and non-linear modes, we first assume that at
$z=0.5$, the linear modes have $k<k^{0.5}_{\rm max}=0.1 h\,{\rm Mpc}^{-1}$. We 
then compute the variance 
$\sigma^2(R^{0.5}_{\rm min}, z=0.5)=\frac{1}{2\pi^2}\int k^2P(k)W^2(kR^{0.5}_{\rm min})dk$
smoothed by a spherical top-hat window function $W(x)=3[\sin(x)-x\cos(x)]/x^3$ with the radius 
$R^{0.5}_{\rm min}=2\pi/k^{0.5}_{\rm max}$. We choose the scale of $R^z_{\rm min}$ at all other redshifts so that 
$\sigma^2(R^z_{\rm min}, z)=\sigma^2(R^{0.5}_{\rm min}, z=0.5)$. We obtain $k^z_{\rm max}=2\pi/R^z_{\rm min}$, 
and $l_{\rm max}^z=k^z_{\rm max}D(z)$ for each z.  

After marginalizing over non-linear $\bar b$'s and $P$'s at each $l$, 
we sum the Fisher matrices for linear-regime parameters over all modes with $l_{\rm min}<l<l_{\rm max}$, with $l_{\rm min}=10$ for all redshifts: 
\begin{equation}
{\bf F^{p q}_{Lens}}(l)=\sum_{l=10}^{l_{\rm max}} f_{sky} (2l+1){\bf F^{p q}_{Lens}}(l).
\end{equation}

An example of the final lensing Fisher matrix is shown in the middle panel of Figure~\ref{FisherbPE}. 
While the constraints on $\bar b_i$ and $P_j$ are highly
correlated for $i=j$ , the corrrelations among $\bar b$ and $P$ at
distinct redshifts are very weak, and we can consider the experiment
to give essentially independent results at every redshift bin.

\section{Multi-tracer redshift space distortion}
\label{RSD}
In this section, we review the basic idea of using redshift space distortion (RSD) to measure the growth 
of structure. This will be implemented in a spectroscopic survey. 
For each redshift shell, galaxies will be made into multiple bins 
of their parent halos' masses. Our measurements will include the redshift space 
power spectra of each sub-sample, and the covariance of all those galaxy bins.

Each redshift shell of the spectroscopic survey will have an
independent Fisher matrix.%, so in this section we will suppress the
%latin index denoting redshift. We use greek indices for galaxy bias bins. 

In the linear regime, galaxy overdensity $\delta^s$ seen in redshift
space will be boosted relative to the matter overdensity $\delta$ due to the 
large-scale inflow bulk motion of galaxies. The first-order large-scale peculiar velocity is related through the continuity 
equation to 
the linear growth rate factor $f \equiv \frac{ \partial \ln G }{ \partial \ln a}$. The redshift-space galaxy clustering therefore encodes 
information on the growth of structure. In Fourier space,
\citet{Kaiser87} derives
\begin{equation}
\delta^s_{\alpha}({\bf k}) = (b_{\alpha} + f \mu^2) \delta({\bf k}) + \epsilon_{\alpha},
\label{Kaiser1}
\end{equation}
with $\epsilon_{\alpha}$ the stochastic portion of the galaxy density
with $\langle \epsilon_{\alpha} \delta \rangle = 0$.  For the RSD
analysis we assume a diagonal stochasticity matrix, 
$\langle \epsilon_{\alpha} \epsilon_{\alpha} \rangle =
\delta^K_{\alpha \beta} / n_{\alpha}$  
i.e. noise in distinct galaxy bins is uncorrelated. $b_{\alpha}$ is the bias of the ${\alpha}$th galaxy bin, 
and $\mu$ is the cosine of the angle between the $k$ vector and the
line of sight.

Following BC11, the covariance of the multi-tracer RSD 
measurement is:
\begin{eqnarray}
\label{cij}
C_{\alpha \beta}({\bf k}) & = & {\rm Cov}(\delta^s_{\alpha}({\bf k}),\delta^s_{\beta}({\bf k}))  \\
&=& (b_{\alpha}+f\mu^2)(b_{\beta}+f\mu^2) G^2P_{CMB}({\bf k}) + {\cal E}_{\alpha \beta},  \nonumber \\
{\cal E}_{\alpha \beta} & \equiv & \langle \epsilon_{\alpha} \epsilon_{\beta} \rangle. 
\end{eqnarray}
The free parameters in this measurement are: the biases
$b_{\alpha}$ of the galaxy bins, assumed to be scale 
independent;
the growth rate $f$, the linear mass power spectrum $P=G^2P_{\rm CMB}$, where $G$ is the linear 
growth function and $P_{\rm CMB}$ is the power spectrum at the epoch
of recombination. 
For each mode at each redshift bin, we have the Fisher 
matrix \citep{TTH} of RSD:
\begin{equation}
{\bf F^{p q}_{RSD}}({\bf k}) =\frac{1}{2}{\rm Tr}\left[{\bf C}({\bf k})^{-1}{\bf C}({\bf k})_{,p} {\bf C}({\bf k})^{-1} {\bf C}({\bf k})_{,q}\right]
\end{equation}
where $p, q\in\{f, P, b_1, b_2...b_{N_b}\}$. 
We assume that galaxies are binned by the mass of their parent halos,
use three log mass bins for each decade of mass ($N_b=15$ mass
bins for the case  
of $M_{\rm min}=10^{11}M_{\odot}$). The size of the Fisher matrix is $(N_b+2)\times(N_b+2)$. 
This multi-tracer RSD method improves the constraint of $fG$ 
by a factor of up to 6.4 %$6.4=\sqrt{21/0.5}$, 
compared to the standard RSD method, where all galaxies are placed in
one single bias bin (BC11).
Without any prior knowledge of galaxy bias, neither method can
constrain $f$ or $G$ alone, only the product $fG$.

Nonlinear $k$ 
modes will not be used in our analysis, applying the criteria from
section~\ref{nonlin}. 
We sum over modes within $k_{\rm min}<k<k^z_{\rm max}$, 
where $k_{\rm min}=l_{\rm min}D(z)$, to yield a total RSD Fisher
matrix for our redshift bin:
\begin{equation}
{\bf F^{p q}_{RSD}({\bf k})}=\frac{V}{(2\pi)^3}\sum_{\mu=-1}^{1} \Delta \mu \sum_{k=k_{\rm min}}^{k_{\rm max}} 4\pi k^2 {\bf F^{p q}_{RSD}}(k)\Delta k.
\end{equation}
Here $V$ is the surveyed volume within the redshift bin under consideration.
When integrating the RSD matrix over $\mu$, we are careful to remove a
section around $\mu=0$ representing the 
number of transverse modes used in constructing ${\bf F_{\rm Lens}}$
for the same bin of redshift and $k=l/D$.  This avoids double-counting
the information in the transverse modes of the spectroscopic survey if
we are analyzing overlapping RSD and lensing surveys.  For
non-overlapping surveys, we do not exclude the $\mu=0$ modes from the
RSD information.

\subsection{Combining RSD and lensing Fisher matrices}
Note 
that the RSD Fisher matrix is degenerate in the $f$--$b$ direction.
Combining RSD measurement with a lensing survey can break the degeneracy between 
$b$ and $f$, and hence yield a tighter constraint on $\gamma$. 
There may, however, be non-zero covariance between $b$ or $P$
values at different redshifts in the lensing measurement. So when
combining the 
constraints from lensing with those from RSD, the parameters in each
$z$ bin can not be treated independently. 
We will need to create a large joint Fisher matrix for biases, $f$,
and $G$ at all redshifts, so we
concatenate ${\bf F^{p q}_{RSD}}$ from all 19 redshifts into a
single block-diagonal RSD matrix.  

We have to convert the lensing and RSD Fisher matrices to
encompass a common set of parameters, then sum them, being careful not
to double-count information.  The final Fisher matrix will contain
entries for the $N_b$ bias values $b_{i\alpha}$, plus the growth and
growth rate $G_i$ and $f_i$ at each of the 19 measurable redshift
bins, giving a final dimension of $19\times (N_b+2)$.  The matrix is
nearly block-diagonal with isolated redshift blocks, because ${\bf
  F_{\rm RSD}}$ is completely decoupled between redshift bins, and ${\bf
  F_{\rm Lens}}$ is nearly so.  We retain the full matrix, however,
for completeness.
 
The lensing Fisher matrix elements for galaxy bias refer to the
weighted mean bias $\bar b_i = \sum_\alpha w_{i\alpha}b_{i\alpha}$ for each redshift. 
We can convert the constraint on the weighted mean bias into a joint
constraint on the individual bins' biases using the known weights $w_{i\alpha}$ from the Appendix.

\section{results}
\label{result}
\begin{figure}
\vspace{-0.8cm}
\resizebox{\hsize}{!}{   
\hspace{-1.7cm}
\includegraphics[angle=0]{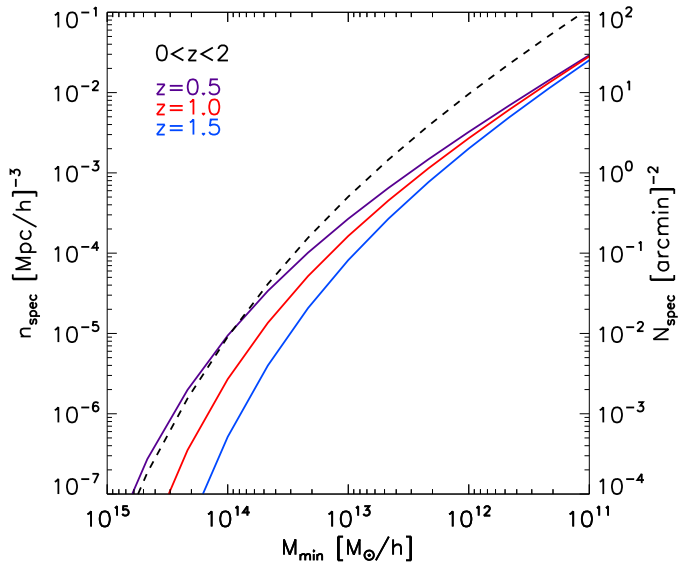}} 
\caption{Galaxy number density $n_{\rm spec}$ in the spectroscopic survey versus the minimum halo mass $M_{\rm min}$ 
at three different redshifts estimated from halo model.  We assume one
spectroscopic target per halo. Black dashed line shows the projected galaxy number density versus $M_{\rm min}$ at $0<z<2$.
}
\label{n_min}
\end{figure}

In this section, we compare the constraints on the growth of structure 
from having a lensing photo-$z$ survey, a spectroscopic redshift survey, and the combination of them. 
We will also investigate the case of having the two surveys over separate volumes. 
We will explore how the results may depend on the depth of the spectroscopic survey and the photo-$z$ survey.
% Since we assume having both surveys covering a fixed area of the sky, $f_{sky}=0.5$, varying the number 
% of spec-$z$ or photo-$z$ galaxies is changing their survey depths. 
For the spec-$z$ sample, we 
usually label the survey depth as the minimal halo mass $M_{\rm min}$,
since we assume that the spectroscopic targets are the central
galaxies of all halos with $M>M_{\rm min}.$ Figure~\ref{n_min} plots
the space density of targets vs $M_{\rm min}$ at a few nominal
redshifts, plus the total projected sky density of targets vs $M_{\rm min}$.
For example, 
$M_{\rm min} \sim 10^{12}h^{-1} M_{\odot}$ corresponds to galaxies of Milky Way size or larger, 
with a space density of $\sim 10^{-2.5} h^3 {\rm Mpc}^{-3}$ and sky
density $\sim10\,{\rm arcmin}^{-2}$; having $M_{\rm min} \sim 10^{13} M_{\odot}$ is 
like a Luminous Red Galaxy (LRG) sample,
with a space density of $\sim 10^{-3.5} h^3 {\rm Mpc}^{-3}$ and sky
density $\sim1\,{\rm arcmin}^{-2}$; and $M_{\rm min}=10^{14}h^{-1}
M_\odot$ is a rich cluster survey, with a space density of $\sim
10^{-5} h^3 {\rm Mpc}^{-3}$ at $z<1$ and sky
density $\sim0.01\,{\rm arcmin}^{-2}$.  Keep in mind that the survey
targeting $M>M_{\rm min}$ generally yields the best possible
cosmological constraints for a given target density.

We will assume that the primordial CMB power spectrum is known exactly, unless specified otherwise. We will show that 
for most cases, knowing $P_{\rm CMB}$ to $0.5\%$ gives about the same
growth constraints as fixing it.
\begin{figure*}
\resizebox{\hsize}{!}{  
\hspace{-1.8cm}                                                                        
\includegraphics[angle=0]{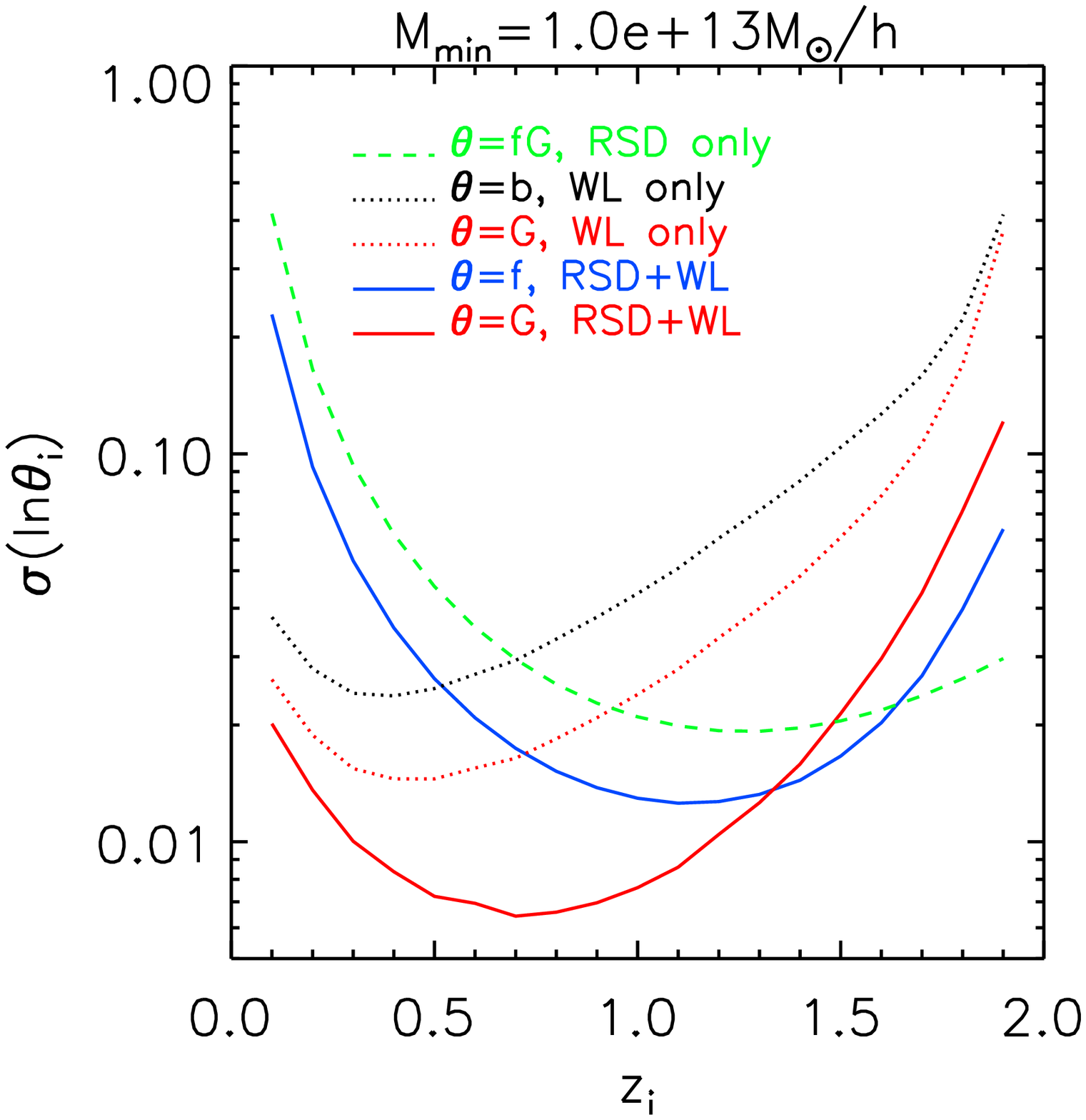}
\hspace{-1.8cm}
\includegraphics[angle=0]{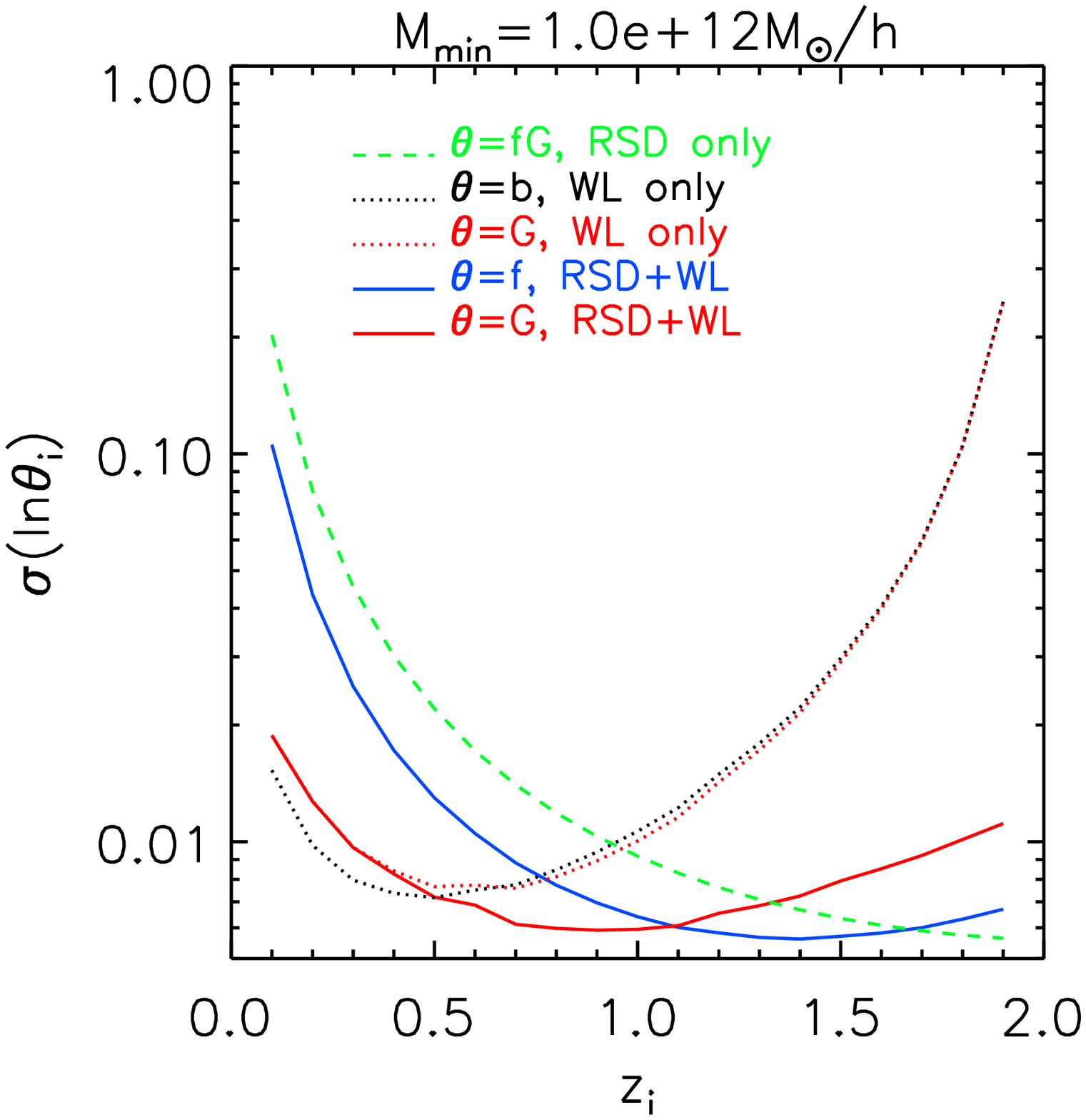}} 
\caption{
Lensing constraints on $\bar b$ and $G$ (dotted lines), RSD constraint on $fG$ (dashed lines) and lensing+RSD 
constraint on $f$ and $G$ (solid lines) in each redshift bin of width
$\Delta z=0.1$ are plotted vs redshift,
for the case of $N_{\rm lens}=10\,{\rm arcmin}^{-2}$. 
{\it Left:} $M_{\rm min} \sim 10^{13} M_{\odot}$ and 
{\it Right:} deeper spectroscopic survey, $M_{\rm min} \sim 10^{12} M_{\odot}$. 
As the depth of spectroscopic survey increase, the constraints on $\bar b$ 
and $G$ from the lensing-galaxy cross-correlation improves, because
the stochasticity of the galaxy sample with respect to mass goes down when smaller halos are mapped. The constraint 
on $fG$ also improves with the spectroscopic survey depth. The joint
constraint on $f$ and $G$ also improves with spectroscopic depth 
at high $z$ in particular.   Adding lensing data to RSD splits the
$fG$ constraint into separate $f$ and $G$ constraints, which are
substantially more precise.}             
\label{dfdPdb1}
\end{figure*}
\subsection{Lensing constraint on $\bar b$ and $G$}
\label{bpConstraint}
We first examine the constraints on $\bar b$ and $P$ at different $z$
bins from ${\bf F_{\rm Lens}}$, the joint analysis of lensing and galaxy density surveys in
purely transverse modes.
Assuming the primordial power spectrum $P_{\rm CMB}$ is known, measuring $P$ is the same as measuring the linear 
growth function $G$. 

The dotted lines in Figure~\ref{dfdPdb1} plot the Fisher
uncertainties in $\bar b_i$ and $G_i$ (the equivalent of $\sqrt{P_i}$) vs
redshift $z_i$.  Each plotted point gives errors after marginalization
over all other parameters. 
We find the measurements of galaxy mean bias and $G$ 
reach percent-level accuracy over a large range of redshifts for
$M_{\rm min}=10^{13}h^{-1}M_\odot$ (left panel), and sub-percent accuracy when
galaxy stochasticity is lower with $M_{\rm min}=10^{12}h^{-1}M_\odot$ (right panel).
The constraint is better at low redshift, easily understood since
higher-redshift lenses have fewer background galaxies to lens and
hence higher effective shape noise in the lensing measurement.

The number of available linear modes increases rapidly at higher $z$,
which should in principle cause constraints to improve with redshift.
While we see this behavior at $z<0.5$, the constraints become weaker
at $z>0.5$, indicating that the increasing shape noise and galaxy
stochasticity dominate the improving mode counts.

\subsection{RSD constraint on $fG$}
Having RSD measurement alone, one can measure the parameter $fG$ after
marginalizing over $f/b$.  The green dashed lines in 
Figure~\ref{dfdPdb1} show the constraint on $fG$ using the multi-tracer RSD method. We find $fG$ is better constrained at 
high $z$, opposite to the lensing constraints on $\bar b$ and $G$ shown in the previous subsection.
 The gain at high $z$ for $\sigma(fG)$ mainly comes from having more
modes as the survey volume $dV/dz$ grows with redshift.
When $M_{\rm min}$ is smaller (comparing the right-hand panel to the left), $\sigma(fG)$ also drops, as expected, 
since we have more galaxies with a broader range of biases. 

BC11 show, and we confirm here, that there is little change in the
cosmological constraints from the multi-tracer RSD analysis from allowing the shot noise level to be a
free parameter instead of fixing the Poisson value.  

\subsection{Combined constraint on $f$ and $G$}
The $\bar b$ and $G$ measurement from lensing-galaxy cross-correlation in transverse modes can be added to 
the RSD analysis in the 3-D spectroscopic redshift survey over the same volume. This will help to break the 
$f$--$b$ degeneracy existing in the case when RSD alone is available. Separate constraints on $f$ 
and $G$ are then achieved after marginalization over all the bias
parameters, leaving 19 $f$'s and 19 $P$'s in the Fisher matrix.
An example of ${\bf\tilde F^{f_i P_j}_{Lens+RSD}}$ is shown in the right hand panel of Figure~\ref{FisherbPE}. 
As expected, the $f-P$ Fisher matrix exhibits very little correlation
between different redshifts.

The solid lines of Figure~\ref{dfdPdb1} plot example constraints
on $f$ and $G$ vs $z$ after marginalizing over all other parameters. 
The redshift dependence of $\sigma(f)$ is similar 
to that of $\sigma(fG)$, and the combined constraint on $G$ improves significantly over the case when lensing 
alone is available. This improvement is more prominent at high $z$. 

\subsection{Constraint on gravity}
\begin{figure*}
\begin{center}
\vspace{0.5cm}
\resizebox{\hsize}{!}{
\hspace{-1.8cm}
\includegraphics[angle=0]{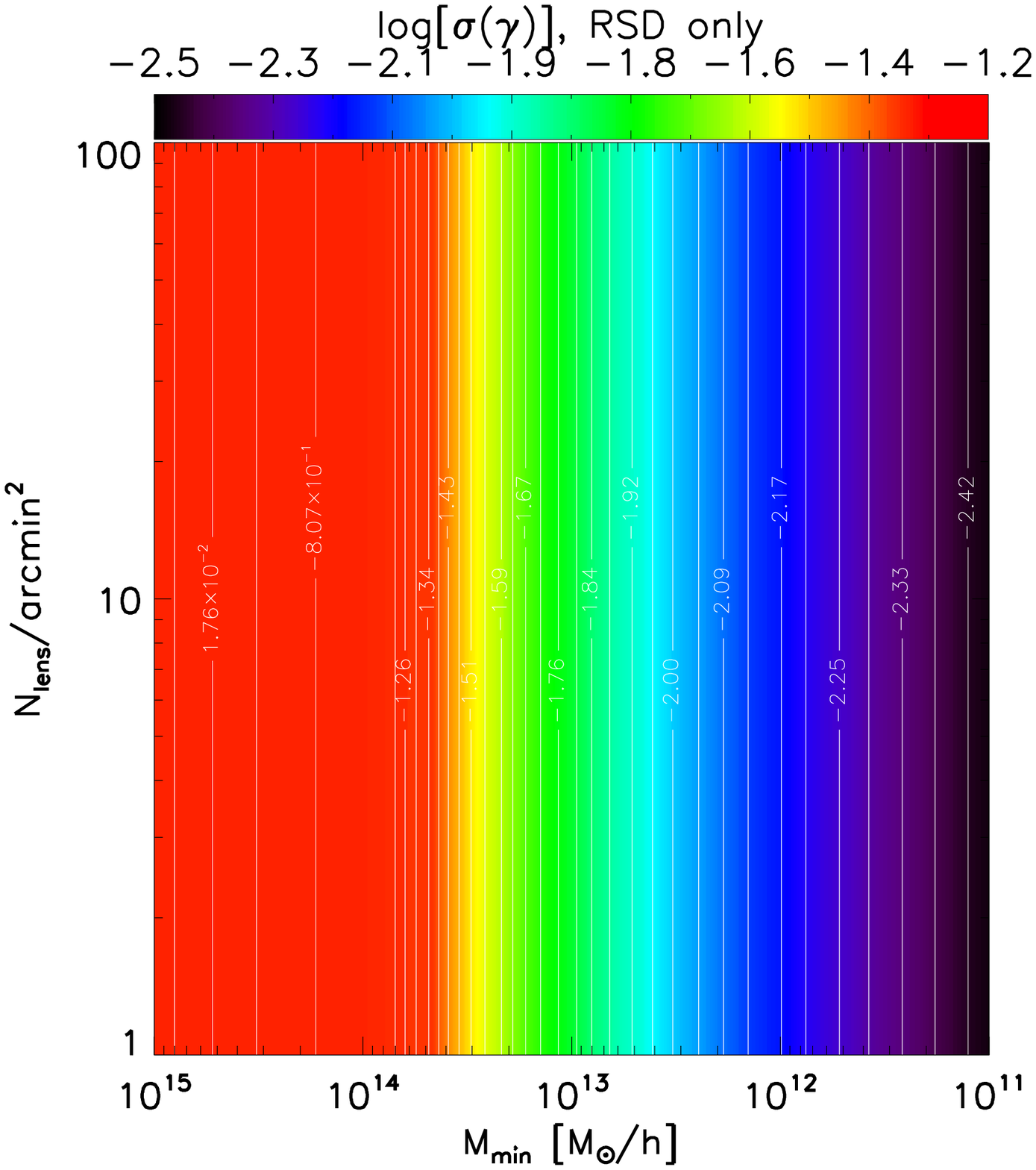}
\hspace{-1.8cm}
\includegraphics[angle=0]{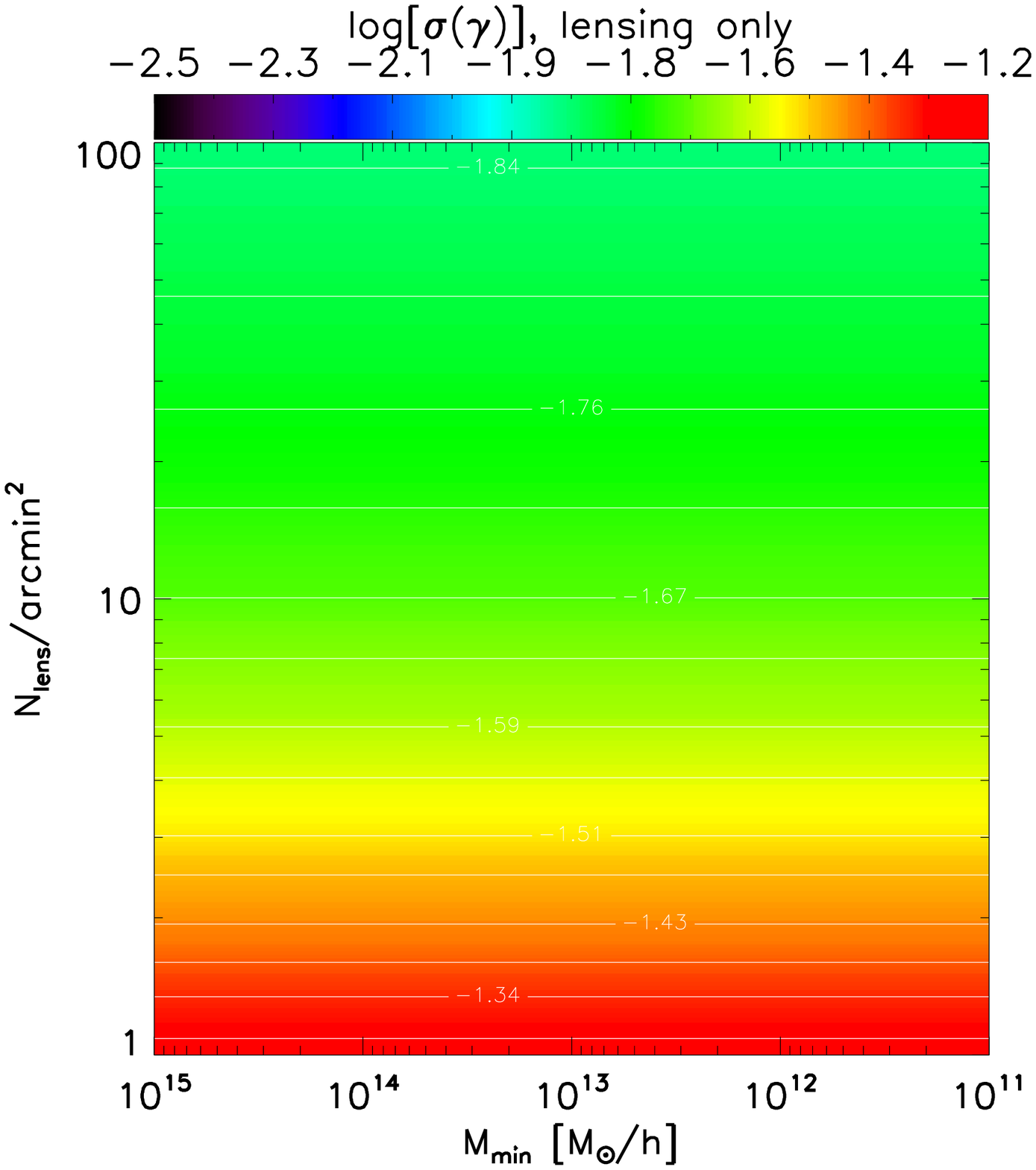}}
\ \ \ \ \ \\
\ \ \ \
\resizebox{\hsize}{!}{
\hspace{-1.8cm}
\includegraphics[angle=0]{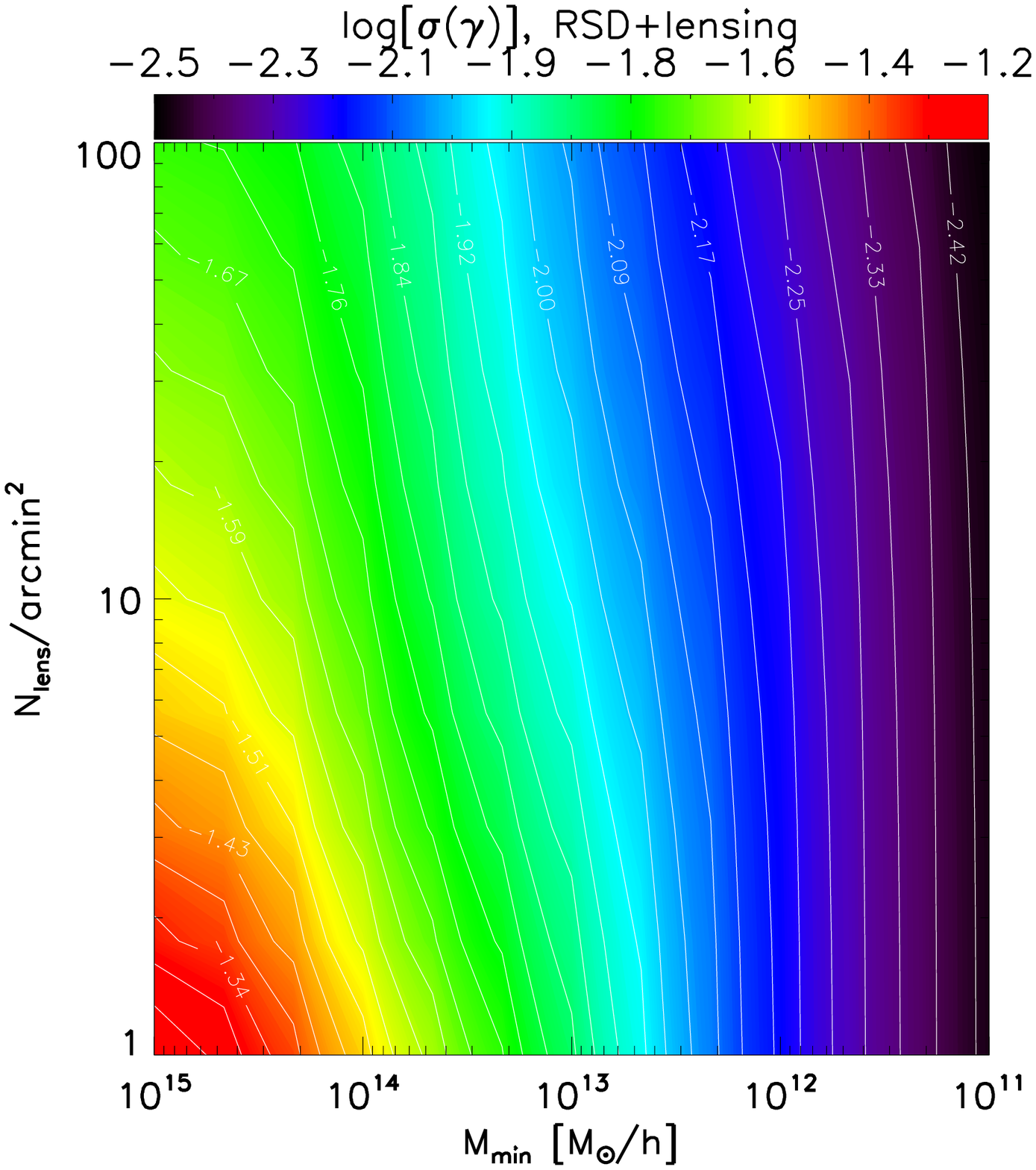}
\hspace{-1.8cm}
\includegraphics[angle=0]{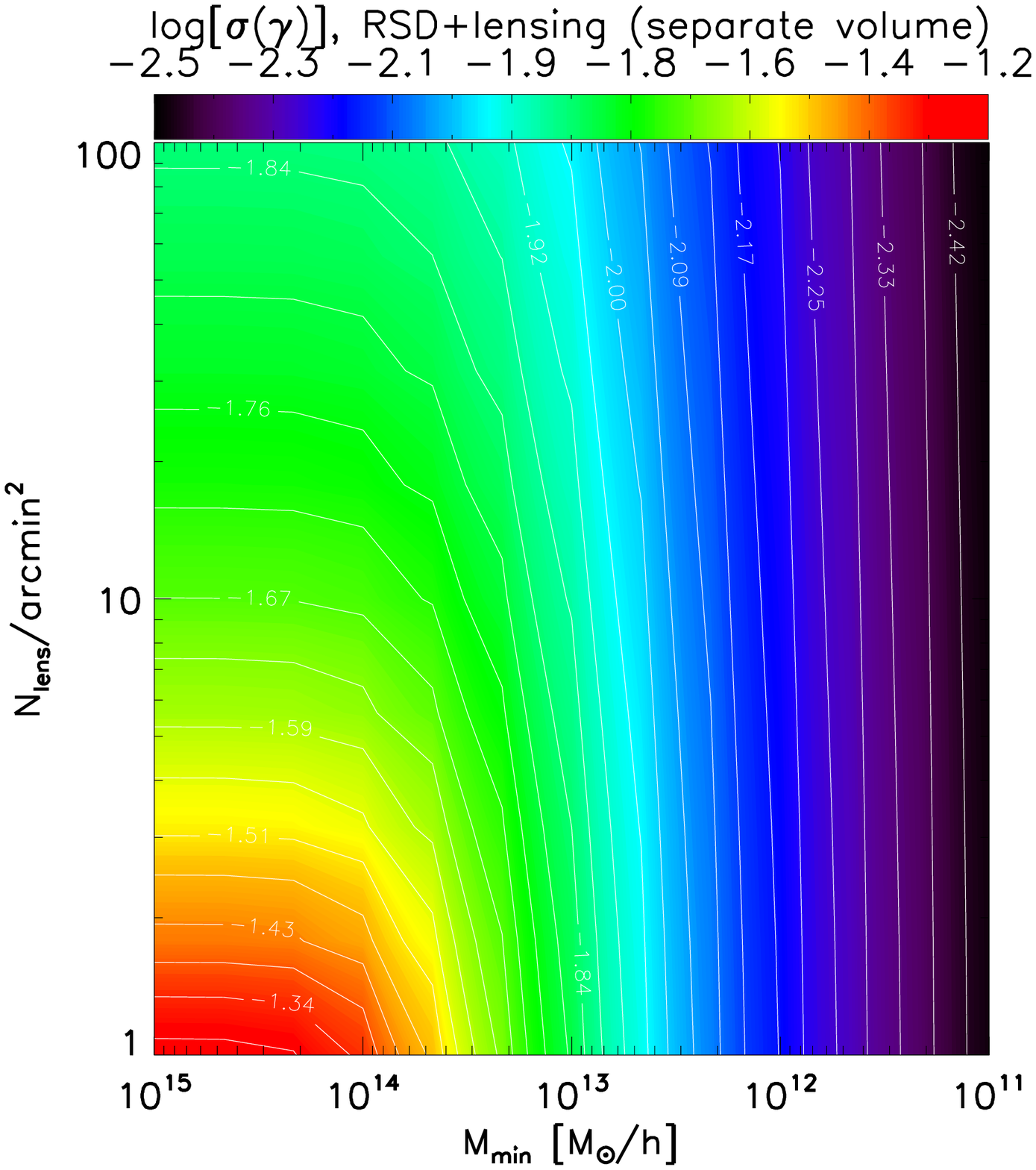}}
\caption{Log of uncertainty $\sigma(\gamma)$ on growth parameter
  $\gamma$  as a function of spectroscopic survey limit $M_{\rm min}$
  and lensing source density $N_{\rm lens}$.  Color scale and contour
  levels are identical for all panels:
 {\it Upper left:} RSD measurement only;
{\it Upper right:} lensing tomography only (including cross-correlation
with photo-$z$ galaxy samples); {\it Lower left:} lensing survey $+$
galaxy redshift survey over the same volume at $0<z<2$, including
cross-correlation between spectroscopic sample and lensing;  
{\it Lower right:}  the same as lower-left panel, but the two surveys
are {\em not} overlapping, so there is lensing cross-correlation with
photo-$z$ galaxies but not the spectroscopic sample. Note that in this case, the total
number of modes are more than the case shown at lower left, i.e. the transverse modes
of the two surveys are now independent.}
\label{contour}
\end{center}
\end{figure*}

\begin{figure*}
\begin{center}
\vspace{0.5cm}
\resizebox{\hsize}{!}{
\hspace{-1.8cm}
\includegraphics[angle=0]{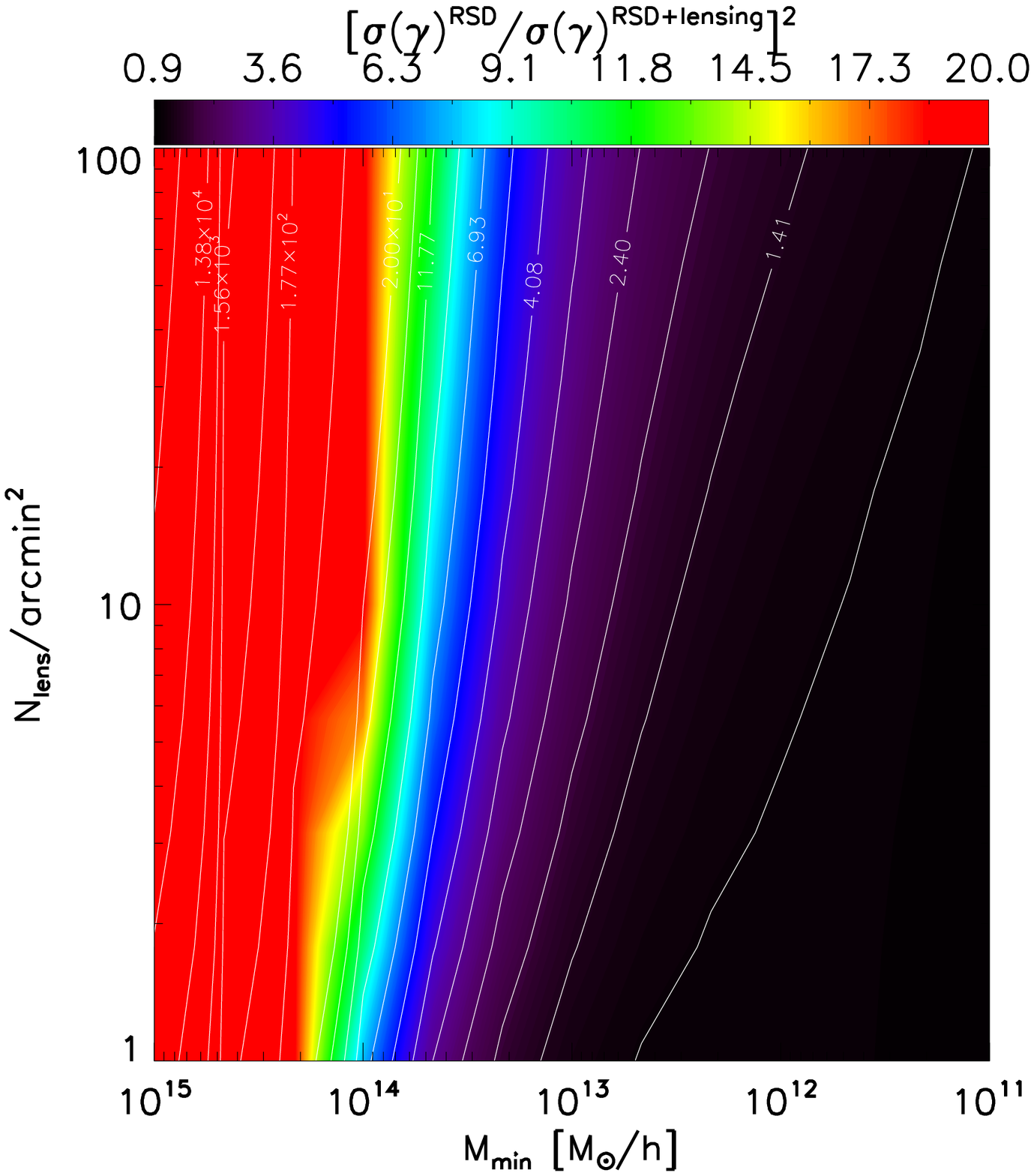}
\hspace{-1.8cm}
\includegraphics[angle=0]{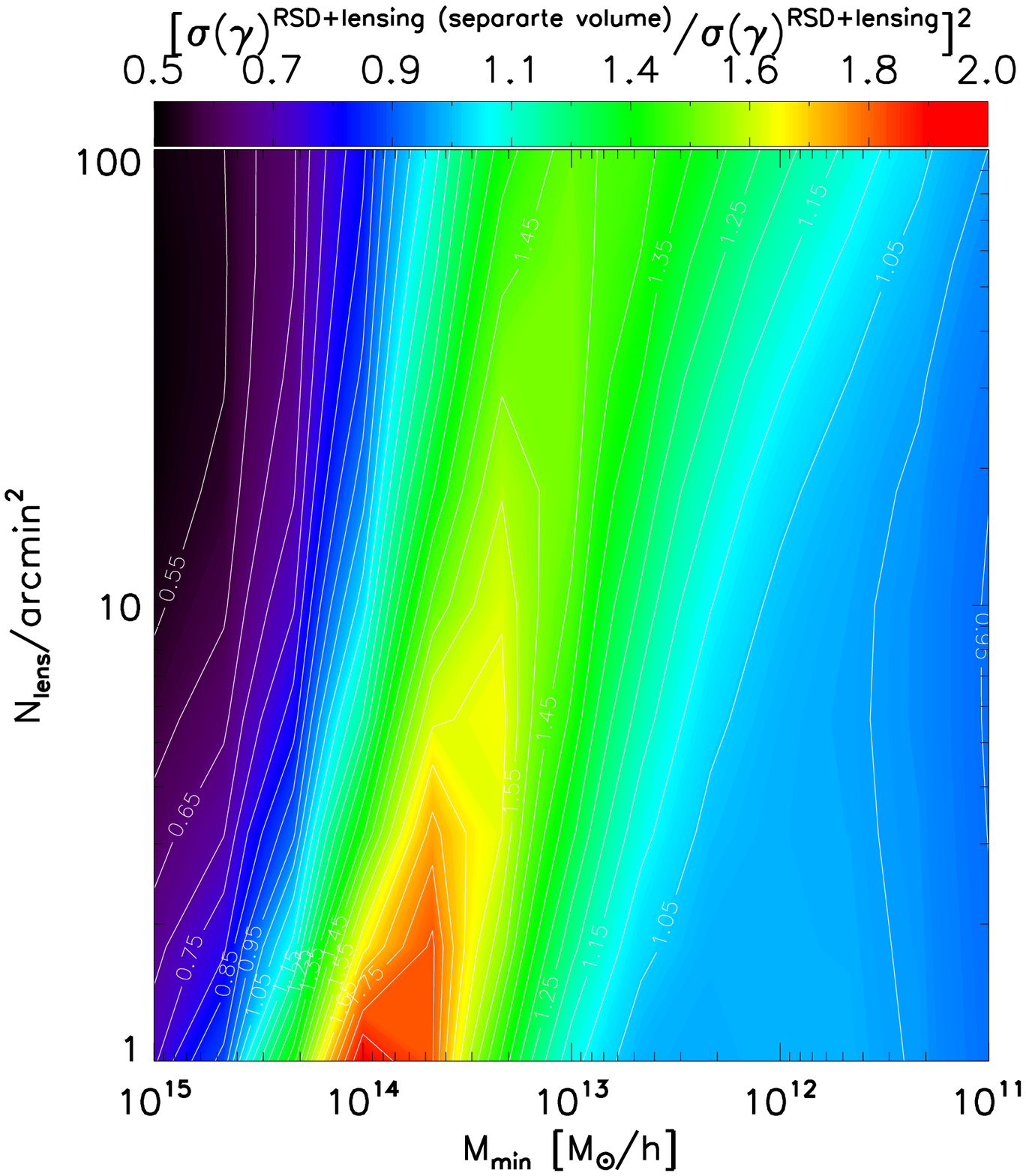}}
\caption{
{\it Left:} the square of the ratios of $\sigma(\gamma)$ from RSD measurement versus that from
lensing+RSD measurement. This is essentially showing how much larger a
survey area the RSD measurement
would need to achieve the same accuracy on $\gamma$ as when lensing
information is added.
{\it Right:} the same as the left but showing the ratio of
lensing+RSD from separate volumes versus lensing+RSD over a common
volume.  The overlapping survey is most beneficial along a band with
$M_{\rm min}\approx 10^{14}M_\odot$.  For deeper spectroscopic
surveys, the separate surveys are nearly equivalent to overlapping, because the
lensing information is a weak addition whether or not overlapping.
Likewise for very shallow spectroscopic surveys and deep lensing
survey (upper left corner), the lensing information dominates and
overlap of RSD is irrelevant.}
\label{ratio}
\end{center}
\end{figure*}

\begin{figure*}
\begin{center}
\resizebox{\hsize}{!}{
\hspace{-1.8cm}
\includegraphics[angle=0]{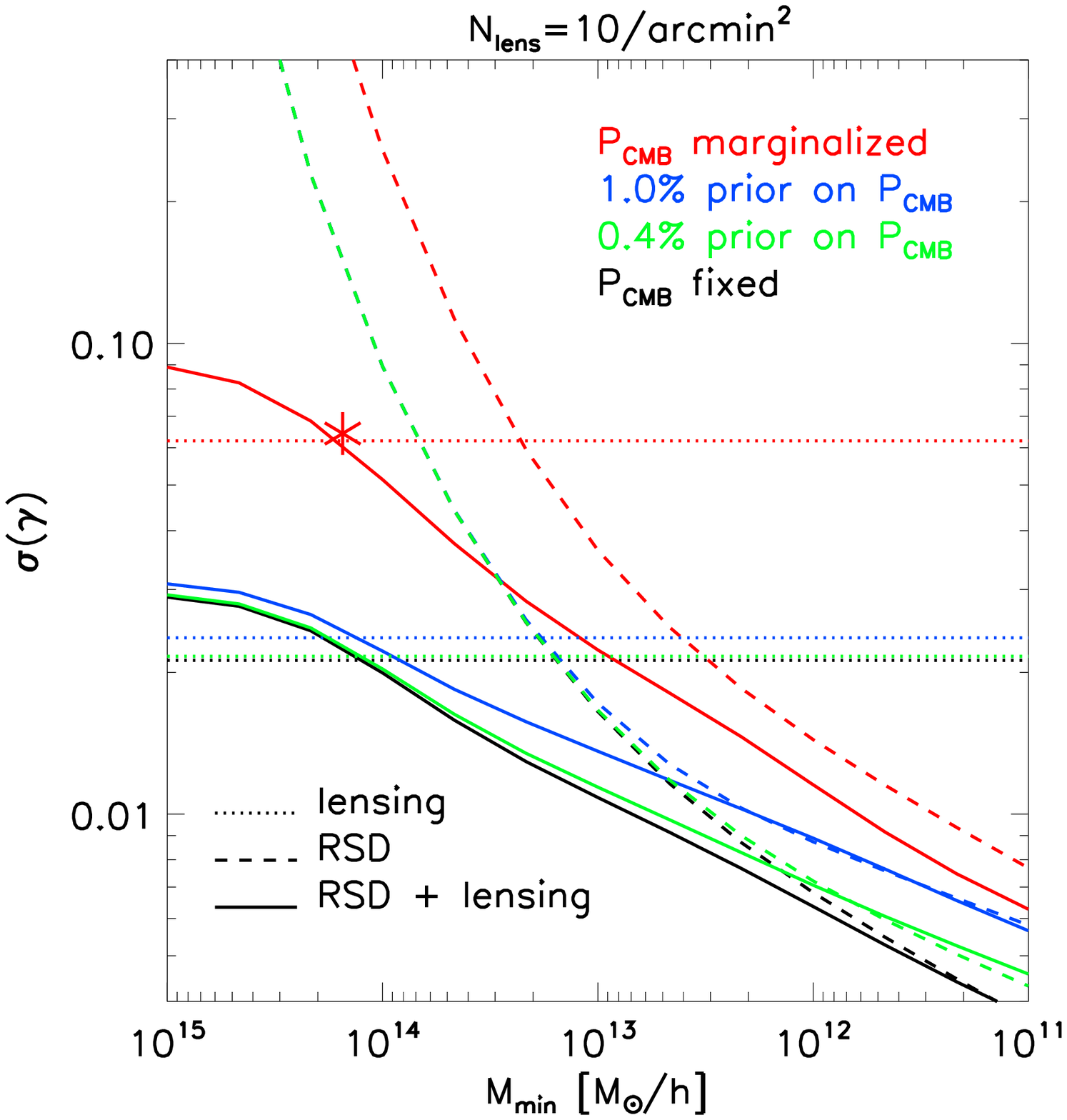}  
\hspace{-1.8cm}
\includegraphics[angle=0]{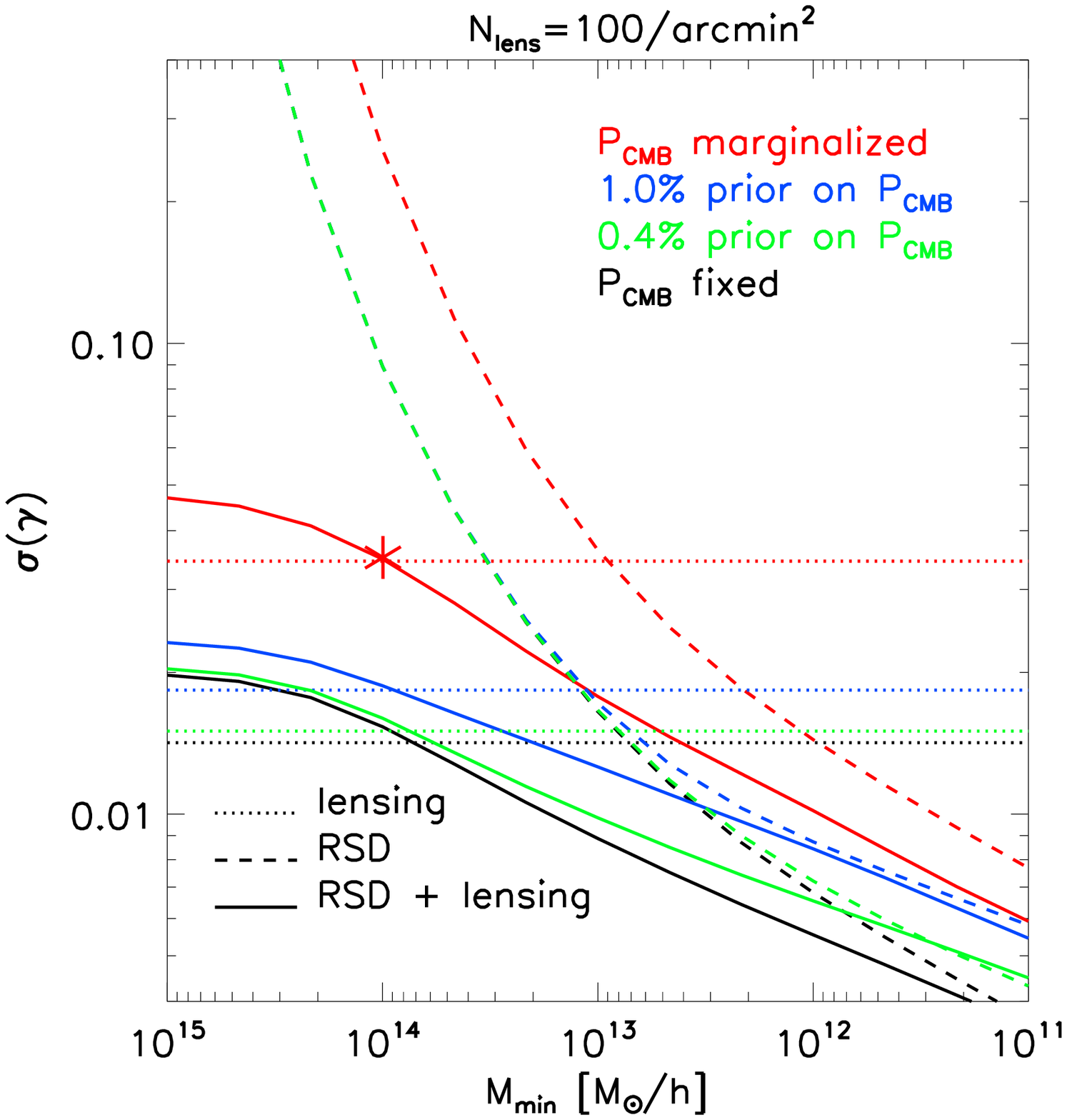}}
\caption{
Comparing $\sigma(\gamma)$ as a function of $M_{\rm min}$ from lensing measurement (dotted lines), RSD measurement 
(dashed lines) and RSD+Lensing (solid lines) with different priors on
$P_{\rm CMB}$ in different colors. {\it Left:} the number density of
galaxies in the lensing survey is $N_{\rm lens}=10\,{\rm
  arcmin}^{-2}$. {\it Right:} ultra-deep lensing survey wtih $N_{\rm
  lens}=100/,{\rm arcmin}^{-2}$. 
When $M_{\rm min}$ is large so the number density of halos in the galaxy redshift survey is small, 
constraints on $\gamma$ mainly come from lensing tomography. When the
spectroscopic survey is deep, addition of lensing data does not improve the 
constraint on $\gamma$ substantially. If we had forecast a
single-tracer RSD analysis instead of multi-tracer RSD, then the gain
from adding lensing to a deep spectroscopic survey would be
larger. Note that adding $0.5\%$ prior on $P_{\rm CMB}$ improves the
constraint on $\gamma$ by a factor 2, and is close to the case of knowing $P_{\rm CMB}$ perfectly.
The red stars indicate where RSD+Lensing is worse than lensing alone in the constraint 
of $\gamma$. This is because in the lensing alone case, we are using the projected 
galaxy-galaxy clustering from the lensing photo-$z$ sample. The stochasticity in this 
case can be lower than that of the spec-$z$ sample, when $M_{\rm min}$ is very large.}
\label{gammaMmin}
\end{center}
\end{figure*}
In the fiducial ${\rm \Lambda}$CDM cosmology, 
a change in $\gamma$ results in predictable changes in the growth history ($f$ \& $G$). Therefore, 
constraints on growth can be translated into a constraint on
$\gamma$. We present $\sigma(\gamma)$ as a 
function of $N_{\rm lens}$, the number density of a lensing survey, and $M_{\rm min}$, 
which is equivalent to the survey depth of 
a spectroscopic redshift survey. In Figure~\ref{contour}, we compare four different cases: 
having RSD alone (I) (top left), lensing alone (II) (top right), RSD
plus lensing over the same volume (III) (bottom left) 
and RSD plus lensing in separate volume (IV) (bottom right).  All results we show assume that the primordial CMB power 
spectrum is known, unless specified otherwise. 
Figure~\ref{gammaMmin} plots $\sigma_\gamma$ vs $M_{\rm min}$ at two
distinct $N_{\rm lens}$ values, for different strengths of prior
knowledge of the amplitude of $P_{\rm CMB}$.  
Marginalization over the normalization of $P_{\rm CMB}$ leads to
$\sigma_\gamma$ constraints about a factor 2 worse than knowing
$P_{\rm CMB}$ exactly, but a prior with 0.5\% accuracy on $P_{\rm
  CMB}$ recovers almost all of this loss.

\subsubsection{(I) RSD alone}
When using multi-tracer RSD from a spectroscopic redshift survey, the result 
strongly depends on the survey depth, or $M_{\rm min}$ (upper left
plot of Figure~\ref{contour} and dashed lines in
Figures~\ref{gammaMmin}. By surveying halos with down to
$M_{\rm min} \sim 10^{12}h^{-1} M_\odot$, RSD alone can already constrain $\gamma$ to $\sim 1\%$. The 
current GAMA survey reaches this survey depth \citep{GAMA11} up to $z\sim 0.5$, 
 but one need to expand the survey 
to cover half of the sky in order to achieve this 
accuracy. As $M_{\rm min}$ increases, $\sigma(\gamma)$ increases rapidly---$\sigma(\gamma) \sim 2\%$ for 
$M_{\rm min}\sim 10^{13}h^{-1} M_\odot$ and $\sigma(\gamma) \sim 10\%$ for $M_{\rm min} \sim 10^{14}h^{-1} M_\odot$ 
(equivalent to galaxy number density of $10^{-6}$--$10^{-5}\,{\rm Mpc}^{-3}$). 

Central LRGs are considered as good samples for RSD measurement, as they reside at the center of 
their host halos therefore should be free from the non-linear finger-of-God effect \citep{OkumuraJing,Hikage11}. 
They are hosted by halos with $M_{\rm min}>10^{13}h^{-1} M_\odot$ \citep[e.g.][]{Zheng09}.
LRGs are, however, an incomplete sampling of halos near
$10^{13}h^{-1} M_\odot$.  Therefore, RSD with LRG samples may not be
as powerful as the $M_{\rm min}=10^{13}h^{-1} M_\odot$ forecast here.

\subsubsection{(II) lensing alone}
When a lensing survey alone is available, we can still estimate the
mass density in the transverse modes using photo-$z$ galaxies.
In this case, the stochasticity between the galaxy density field and the mass 
field may be larger, so we conservatively assume $E_{\rm fid}=0.5$ and add a weak prior on the $\noise$ term 
as we have discussed in section \ref{cross-correlation}. 

For this case of joint shear/density tomography without spectra,
$\sigma(\gamma)$ varies by
a factor of $\approx 5$ when $N_{\rm lens}$ varies from 1 to 100
(Figure~\ref{contour}, upper right). The upcoming DES 
is expected to have $N_{\rm lens}\sim 10\,{\rm arcmin}^{-2}$, which if
scaled to $f_{\rm sky}=0.5$ constrains $\gamma$ down 
to $\approx 2\%$ (Figure~\ref{gammaMmin}, left). Future surveys like
  LSST or Euclid, attaining $N_{\rm lens}\sim 
40\,{\rm arcmin}^{-2}$ in the most optimistic scenario, yield
$\approx 1.25\times$ reduction in $\sigma_\gamma$. 

Removing the CMB constraint on the amplitude of the primordial power spectrum 
may increase the error in $\gamma$ by a factor of 2.

\subsubsection{(III) RSD + lensing (same volume)}
Combining the two surveys will in general help to improve the
constraint on $\gamma$. The $\sigma(\gamma)$ for the overlapping
surveys is plotted in the lower left of Figure~\ref{contour}.  The
left-hand plot of Figure~\ref{ratio} quantifies the improvement from
adding lensing to the spectroscopic survey as the inverse square of
the improvement in $\sigma(\gamma)$, which is equivalent to asking
what factor more survey area the RSD survey would require in order to
match the improvement from the addition of lensing data.
The amount of improvement depends on many factors. 

When the spectroscopic redshift survey is very deep, i.e. 
$M_{\rm min}\sim 10^{12}h^{-1} M_\odot$, RSD alone can already measure 
$\gamma$ at sub-percent level, if combined with $0.5\%$ constraint on
the primordial power spectrum from the CMB (Figure~\ref{gammaMmin}).  The number density of
halo 
redshifts in such a survey is
$n_{spec}\sim 10^{-3}h^3\,{\rm Mpc}^{-3}$, requiring a total of $\sim
10^8$ redshifts over half of the sky.  In this regime, the improvement in the constraint of $\gamma$ by adding 
lensing data is very minor, and changing the lensing survey depth will not affect the result. 
This is consistent with the result of BC11 (see their Figure~3).

Figure~\ref{dfdPdb1} displays dramatic gains in constraint of $f$ and
$G$ at $z>1.2$ when $M_{\rm min}$ is reduced from $10^{13}$ to
$10^{12}h^{-1}M_\odot$, yet only modest gains in $\sigma(\gamma)$ are
seen in Figure~\ref{gammaMmin} or on the left of Figure~\ref{ratio}.  
This is because both the absolute values of $\partial P/\partial \gamma$ and 
$\partial f/\partial \gamma$ become smaller at high $z$, where
$\Omega_m$ is close to 1, so measures of $f$ and $P$ at $z>1$ are less
valuable in constraining gravity under this parameterization.

When only halos with $M>10^{13}h^{-1} M_\odot$ are targeted in the
spectroscopy survey, the benefit of combining with a lensing survey becomes 
more prominent: equivalent to a factor of 2 to 3 increase in survey volume at 
$M_{\rm min}\sim 3\times 10^{13}h^{-1} M_\odot$, 
and more than $10\times$ when $M_{\rm min}\sim 10^{14}h^{-1} M_\odot$!
(See left of Figure~\ref{ratio} and Figures~\ref{gammaMmin}.)
However, even in this regime, lensing is never completely dominant in
the range of $N_{\rm lens}$ we are considering, in the sense that
Figure~\ref{contour} shows that improving the spectroscopic survey
depth is always substantially beneficial for measuring $\gamma$.

\subsubsection{(IV) RSD + lensing (separate volume)}
Here we compare the power of lensing and RSD surveys conducted over a
shared $f_{\rm sky}=0.5$ to surveys that do {\em not} overlap,
covering distinct volumes.
Having two surveys in separate volumes has the 
advantage of having twice as many transverse modes as the case of
overlapping survey volumes; is this advantage outweighed by knowing
the bias of the spectroscopic survey galaxies through the overlapped
lensing survey?

To forecast the $\gamma$ constraints from separate surveys, we make
the following alterations to the Fisher methodology for the combined
surveys:  first, we construct ${\bf F_{\rm Lens}}$ under the
assumption that photo-$z$ samples are being used for the galaxy
density map ($\Rightarrow E_{\rm fid}=0.5$).  Then we marginalize the
$\bar b_i$ values in ${\bf F_{\rm Lens}}$ to leave constraints over
only the $P_i$.  For the RSD Fisher matrix, we marginalize over all
the $b_{i\alpha}$ since no lensing constraints are available, leaving
behind only a constraint on the product $fP$ at each redshift.  We
also allow the RSD analysis to use all transverse modes, since these
are no longer redundant with those in the lensing survey.  The lensing
and RSD Fisher matrices can again be summed and projected onto a
single $\sigma(\gamma)$, plotted in the lower right of
Figure~\ref{contour}.  We also plot, on the right-hand side of
Figure~\ref{ratio}, the effective area gain of the overlapping survey
relative to separate surveys.  Note that this area ``gain'' could be
as low as 0.5, i.e. a loss, since the combined survey does cover only
half the volume of the separate ones.

We find that having two surveys over the same volume is better than
having them separated except for extremely deep lensing or
spectroscopic surveys. The 
improvement is equivalent to a factor of 1.5 to 2 in survey volume when 
$10^{13}h^{-1} M_\odot<M_{\rm min}<10^{14}h^{-1} M_\odot$, but very minor when the spectroscopic redshift survey gets deeper. 
This is true when the primordial CMB power spectrum is known to better than $0.5\%$. If we do not 
employ any CMB constraint and marginalize over $P_{\rm CMB}$, then the $\gamma$ constraint will be degraded for each case, 
but the gain of having overlapping survey volume versus separate volume is larger, e.g. a factor of 3--4 in the regime 
when $M_{\rm min}$ is large. The area gain factor is $\approx1.5$ even when the spec-$z$ is deep. Therefore, having a weak CMB prior 
makes the idea of combining two surveys over the same volume more useful, while a strong CMB prior help to reduce 
$\sigma(\gamma)$ in both cases and narrows the difference between them.

Notice on the lower-right of Figure~\ref{contour} that in the regime
when $M_{\rm min}>10^{13.5}h^{-1}M_\odot$ (shallow spectroscopic
redshift survey), the constraint from lensing measurements is dominant
and the depth of the non-overlapping spectroscopic survey becomes
irrelevant.

In summary, combining two surveys help most, relative to separate surveys, when 
the spectroscopic redshift survey is modestly sparse, $M_{\rm
  min}\approx10^{13.5}h^{-1}M_\odot$, in the range of LRG surveys.
When the spectroscopic survey is deep ($M_{\rm
  min}<10^{13}h^{-1}M_\odot$), it dominates the error budget and it
matters less whether the lensing survey overlaps or not.  On the other
hand, when the spectroscopic survey is very shallow ($M_{\rm
  min}>10^{14.5}h^{-1}M_\odot$), then even a modest lensing survey
($N_{\rm lens} > 5\,{\rm arcmin}^{-2}$) dominates the information, and
it matters less whether the spectroscopic survey is coincident.
There is, however,  {\em no} regime of feasible large-scale surveys for which the
separate surveys constrain $\gamma$ {\em better} than overlapping surveys.

\section{conclusion and discussion}
\label{conclusion}
We have shown from Fisher matrix forecast that the constraint on the growth of structure and gravity 
can be reduced percent-level or even sub-percent level by combining a spectroscopic redshift 
survey with a photo-$z$ weak lensing survey over the same volume.
Whereas BC11 merely assumed that some measure
of galaxy bias was available to add to RSD information, we verify here that a
realistic tomographic weak lensing survey does in fact yield bias
information sufficient to realize a substantial gain in accuracy on
the growth parameter $\gamma$.

Following the suggestions of \citet{Pen04}
we use the shear-galaxy 
cross-correlation to measure the galaxy bias in the transverse
modes---a measurement which is free of sample variance---and apply it
to the multi-tracer RSD analysis in a
spectroscopic redshift survey \citep{MS09,BC2011}.  
The combination of the two surveys make it possible to measure the
linear growth function $G$ separately from its derivative $f=d\ln G /
d\ln a$, whereas RSD alone can only measure the product $fG$.

The performance of multi-tracer RSD measurement depends on the
spectroscopic survey depth, the range of galaxy 
biases in the sample, and the number of linear modes available. The
performance of the shear$+$galaxy analysis on the transverse modes
depends on: (1) the level of stochasticity between the galaxies and the projected mass, (2) the depth of 
the lensing survey, or shape noise. When combining two measurements together over the same volume, 
the results will depend on all those factors that affect each of the survey.

We have demonstrated that for the constraint of the $\gamma$ parameter, 
combining two surveys is better than having each of them alone,
roughly a factor 1.5 improvement (in survey-area terms) in the regime
of likely feasible surveys: source density $N_{\rm lens}\approx
10\,{\rm arcmin}^{-2}$ in the lensing survey, and galaxy surveys
complete for halos in the cluster or small-group range $M_{\rm
  min}=10^{13}$--$10^{14}h^{-1} M_\odot$, similar to LRG surveys.
For $M_{\rm min}>10^{13} h^{-1} M_\odot$, the lensing$+$RSD survey has
constraints many times more powerful than the RSD survey alone.
The $\gamma$ parameterization of growth predicts very little change at
$z>1$; a different model for deviations from General Relativity could
gain even more from the combination of lensing and RSD surveys.

Having prior constraints on the amplitude of the primordial power spectrum from the CMB is useful in general. 
Knowing $P_{\rm CMB}$ to $0.5\%$, easily within the statistical power of
Planck, garners most of the $\approx2\times$ gain in accuracy on
$\gamma$ that is possible with perfect {\it a priori} knowledge of
$P_{\rm CMB}$.  If $P_{\rm CMB}$ is more poorly known, the gain of
having overlapping surveys over the case of separate survey volume is increased.

During preparation of this paper, \citet{Gaztanaga11} released very
similar calculations of the benefit of coincident lensing and
spectroscopic surveys.  Their assumed survey configurations and free
parameterizations differ substantially from ours, so direct
quantitative comparison is not possible.  
In the particular case of constraints on $\gamma$, they find
overlapping surveys reducing $\sigma(\gamma)$ by $\approx 2.4\times$ compared to seperated survey 
volumes, equivalent to a $6\times$ increase in survey area in the
language of our Figure~\ref{ratio} where we find $\approx 1.5\times$
areal gain.  This is qualitatively consistent
with our conclusion, but the origin of the substantial quantitative difference 
is difficult to ascertain given the different assumptions about survey
characteristics.  \citet{Gaztanaga11} also find substantial gains in
accuracy of dark energy equation-of-state determination from
overlapping surveys.  Our analysis holds this fixed so we would not
have detected these gains; we plan to broaden our analysis to the case
of unknown distance-redshift relations in the near future.
 
We notice that bias measurement can in principle also be measured using the same spectroscopic sample from the galaxy 
bispectrum \citep{Simpson11}. If the same accuracy of bias can be obtained in this method as using lensing, 
one can simply use one spectroscopic redshift survey to obtain the same measurement, which might be another attractive 
survey strategy since no lensing survey is needed.  The lensing survey
is, however, a straightforward measure of the galaxy bias, free of
assumptions about perturbation theory, second-order bias, and other
issues with the bispectrum.

Use of smaller-scale modes are attractive in the sense that one may gain many more modes from
the same volume of survey. Growth test statistical accuracy improves rapidly with
increasing $k_{\rm max}$.  Non-linear effects in the density or
velocity field and scale-dependent bias may, however, ruin the attempt to achieve percent-level constraint on parameters.
Efforts have
been made to improve RSD predictions for smaller-scale modes \citep{Scoccimarro04, Jennings11,
Hikage11,Tang11}, e.g. $k_{\rm max}\sim0.3$, though it is important
that predictions be made for galaxies or halos rather than all mass
particles in an $N$-body simulation \citep{Jennings11}, e.g 
see \citet{Reid11} for modeling of halos.  Better
understanding of the non-linear biases of different tracers is
required before one can confidently select the $k_{\rm max}$ that
admits the most accurate growth constraints.

\section*{ACKNOWLEDGEMENT}
This work was supported by NASA grant NNX11AI25G, NSF grant AST-0908027, and DOE grant
DE-FG02-95ER40893.  The authors thank Ravi Sheth, Enrique Gaztanaga,
and Martin Eriksen for assistance and insight. YC thanks Shanghai Astronomical Observatory, 
Purple Mountain Observatory, University of St Andrews, University of Edinburgh and Durham University
for their hospitality during his visit, and Alan Heavens, Catherine Heymans and Andy
Taylor for their useful discussion.

\bibliography{LensingConstraint}
\bibliographystyle{mn2e}

\appendix
\section{optimal weighting and minimal stochasticity}
We take the halo model description of the stochasticity $E$ that has been
developed in CBS. The basic idea is to split halos into different mass bins,
apply the optimal weight $w_{\rm opt}$ to each of them:

\begin{equation}
 \label{halomodelw}
w_{\rm opt}(m) = \frac{m\,u(k|m)}{\bar\rho}
          + F_v\, \frac{v(m)\, P(k)}{1 + (nv^2)_h P(k)},
\end{equation}
where the first term on the right hand is equivalent to mass weighting,
and the second term being close to bias weighting, $v(m)$ is the halo bias
respect to the `continuous halo field', $u(k|m)$ is the Fourier transform
of the NFW halo profile (NFW), $\bar{\rho}$ is the mean mass density.
\begin{eqnarray}
F_v &=& 1 -
        \int_{M_{\rm min}}^\infty dm\,\frac{dn}{dm}\frac{m\,u(k|m)}{\bar\rho}\,v(m), \\
 (nv^2)_h &=& \int_{M_{\rm min}}^\infty dm\,\frac{dn}{dm}\,v^2(m).
\end{eqnarray}
We then obtain the corresponding stochasticity between the weighted halos field
and the mass field:
\begin{equation}
\label{halomodelE}
E_{\rm opt}^2 = 1 - \frac{C_{wm}^2}{C_{mm}C_{ww}} = 1 - \frac{n_w\,C_{wm}}{C_{mm}}
\end{equation}
where
\begin{eqnarray}
 \label{pk}
 C_{ww} &=& v_w^2\, P(k) + \noise_w ,\\
 C_{wm} &=& v_w\, P(k) + \noise_\times ,
\end{eqnarray}

\begin{eqnarray}
 \label{nw}
 v_w &=& \int_{M_{\rm min}}^\infty dm\, \frac{dn}{dm}\, \frac{w(m)}{n_w}\, v(m), \\
 n_w &=& \int_{M_{\rm min}}^\infty dm\, \frac{dn}{dm}\, w(m) ,\\
 \noise_w &=& \int_{M_{\rm min}}^\infty dm\, \frac{dn}{dm}\, \frac{w^2(m)}{n_w^2}, \\
\noise_\times &=& \int_{M_{\rm min}}^\infty dm\, \frac{dn}{dm}\,
                  \frac{m\,u(k|m)}{\bar\rho}\, \frac{w(m)}{n_w}, \\
 \noise_m &=& \int_0^\infty dm\, \frac{dn}{dm}\,
              \frac{m^2\,|u(k|m)|^2}{\bar\rho^2}
\label{noisem}
\end{eqnarray}
This model has been shown to be in good agreement with simulations \citep{CBS11}. We will use
$E_{w_{\rm opt}}$ as the fiducial value of stochasticity in this paper.

\end{document}